\documentclass{aastex}
\usepackage{spr-astr-addons}
\usepackage{url}

\RequirePackage{color}

\begin{document}

\title{Improved SOT (Hinode mission) high resolution solar imaging observations }
\slugcomment{Not to appear in Nonlearned J., 45.}
\shorttitle{Improved SOT high resolution solar imaging observations} \shortauthors{Goodarzi
et al.}

\author{H. Goodarzi\altaffilmark{1,2}} \and \author{S. Koutchmy\altaffilmark{2}}
\email{\h.godarzi@tabrizu.ac.ir}
\and
\author{A. Adjabshirizadeh\altaffilmark{1}}

\altaffiltext{1}{Department~of~Theoretical~Physics~and~Astrophysics, Faculty of Physics, ~University~of~Tabriz,~Tabriz~51664,~Iran.}

\altaffiltext{2}{Institut d'Astrophysique de Paris UMR 7091, CNRS and UPMC (Sorbonne Univ.), 98 Bis Bd Arago, 75014, Paris (France) }

\begin{abstract}

We consider the best today available observations of the Sun free of turbulent Earth atmospheric effects, taken with the Solar Optical Telescope (SOT) onboard the Hinode spacecraft. Both the instrumental smearing and the observed stray light are analyzed in order to improve the resolution. The Point Spread Function (PSF) corresponding to the blue continuum Broadband Filter Imager (BFI) near 450 nm is deduced by analyzing i/ the limb of the Sun and ii/ images taken during the transit of the planet Venus in 2012. A combination of Gaussian and Lorentzian functions is selected to construct a PSF in order to remove both smearing due to the instrumental diffraction effects (PSF core) and the large-angle stray light due to the spiders and central obscuration (wings of the PSF) that are responsible for the parasitic stray light. A Max-likelihood deconvolution procedure based on an optimum number of iterations is discussed. It is applied to several solar field images, including the granulation near the limb. The normal non-magnetic granulation is compared to the abnormal granulation which we call magnetic. A new feature appearing for the first time at the extreme- limb of the disk (the last 100 km) is discussed in the context of the definition of the solar edge and of the solar diameter. A single sunspot is considered in order to illustrate how effectively the restoration works on the sunspot core. A set of 125 consecutive deconvolved images  is assembled in a 45 min long movie illustrating the complexity of the dynamical behavior inside and around the sunspot.
\end{abstract}



\section{Introduction}

A new era in solar surface imaging techniques at high resolution has emerged in recent years thanks to i/ the introduction of adaptive optics and computer aided speckle imaging methods at ground-based solar observatories with 1 to 1.6 m diameter aperture telescopes; ii/ the availability of a wealth of diffraction limited images from the Solar Optical Telescope (SOT) of the Hinode mission, (Tsuneta et al., 2008; Suematsu et al., 2010) free of turbulent Earth atmospheric smearing effects with perfectly reproducible properties. It is then useful to carefully analyze all instrumental effects influencing these space-borne images in order to extract the finest details of the solar surface up to its extreme limb, in particular for the knowledge of dynamical effects produced by the magnetic fields continually emerging to the surface. This includes the case of very large concentration of magnetic flux as is the case in sunspots, although only the case of the less complicated single, unipolar, well developed sunspot near the center of the Sun will be considered here. We leave for a another paper the main application we want to consider of sunspot evolution, without addressing the question of the origin of the sunspot phenomenon in deep layers of the Sun.

A qualitative improvement in very fine details of the dynamical layer immediately above the photosphere, as observed using an H line emissions of CaII filter, has already been the subject of several Hinode SOT papers devoted to the study of spicules, their roots and their extension towards the corona (see for ex. Tavabi, Koutchmy and Adjabshirzadeh, 2011 \& 2014). We will not apply the rather qualitative method described in Tavabi, Koutchmy and Adjabshirzadeh (2013), in the case of photospheric surface features. Instead, more quantitative methods should be used considering phenomena dominated by radiative transfer aspects, where the optical thickness effects are important, which was not the case for spicules observed off-disk with a broad band H CaII filter (Tavabi, Koutchmy and Adjabshirzadeh, 2011).
Telescopes cannot work in practice at their ultimate diffraction-limited quality due to optical aberration(s) arising from imperfections, thermal changes, dirt and maladjustments from optimum parameters after working for a period of time, including the defocusing. In the case of the Hinode SOT we deal with very small effects.

Scattered light and stray light are mathematically described using the extended wings of the point spread function (PSF), which causes an image of the point source of light to be spread (Johnson, 1972). This effect can be enhanced when long averaging exposure times are used in ground based observations. In ideal conditions of a non-obstructed pupil, the PSF of a diffraction limited telescope is described by the Airy pattern. Usually optical telescopes work in situations that are far from these ideal conditions. Recall that in the case of a ground based instrument the earth's atmosphere will limit observations. The access to seeing-free data should be of great benefit with space-born instruments. For space-born telescopes, the PSF was tentatively determined in laboratory conditions before the flight (Suematsu et al., 2010). However, after the launch some in-flight adjustments are performed and because the whole instrument works in a different situation in space, it is better to try to measure the PSF "in situ", using indirect methods such as the Mercury transit, solar eclipses, or using the limb of the Sun (see Mathew et al., 2009; Wedemeyer-B\"{o}hm, 2008; Wedemeyer-B\"{o}hm and Rouppe van der Voort, 2009), for earlier deductions of the PSF of the Hinode SOT. Note that its optical telescope assembly (OTA) has a 50 cm outer diameter with three spiders (4 cm wide) holding the 17.2 cm central mirror, producing an obstruction in the path of light (Tsuneta et al., 2008; Suematsu et al., 2008).The diffraction-limited PSF of a circular aperture of 50 cm clear aperture is described by the Airy function where the wavelength is a parameter. For more complicated apertures the PSF can be deduced from the Fourier transform of the pupil shape when it is known. Note that the central obscuration (in linear scale 0.344) is not exactly a part of the pupil because it is well before the main mirror and it is illuminated by the full Sun "from the back", producing some amount of spurious stray light towards the primary mirror that is further reduced by the Gregorian system used in the telescope (see Suematsu et al., 2008). The theoretical diffraction-limited PSF with three extended spiders was discussed in Wedemeyer-B\"{o}hm and Rouppe van der Voort (2009) in terms of the modulation transfer function (MTF), which is the Fourier transform of the PSF.

It is not easy to apply the concept to image restoration using the "theoretical" MTF because the effect of the noise dominates at high spatial frequencies and should be taken into account. The Wiener 2D filter (see Koutchmy and Koutchmy, 1989 and Bracewell, 1976) is a statistically justified solution but it does not take into account the local conditions with a position-dependent noise. Wedemeyer-B\"{o}hm (2008) utilized an eclipse and the Mercury transit to evaluate the PSF of the SOT by qualitatively introducing a convolution of an "ideal" PSF part (Airy function) and a "non-ideal" PSF part (overall, described by a Voigt function). Note that the edge of the Moon is far from being smooth, with the profile produced by mountains and craters to be reconstructed for the time of observation because of the lunar libration and this is not easy to introduce for deducing the edge-like smearing function. Recent lunar profiles reconstructed from the altimetry data of the Kaguya mission (see Koutchmy et al., 2013) can now be introduced, but it does not seem to have been used to deduce the PSF of the SOT of Hinode. Mathew et al. (2009) determined the PSF of the SOT also by analyzing the transit of Mercury. They introduced a combination of four Gaussians with different widths and weights in order to model the PSF. Gaussian functions can describe the narrow core of the PSF, however they cannot easily account for the far wings of the PSF describing the stray light (see for ex. Koutchmy \& Koutchmy, 1974; Koutchmy, Koutchmy \& Kotov, 1979 and the recent papers by Wedemeyer-B\"{o}hm 2008; DeForest, Martens \& Wills-Davey, 2009; Wedemeyer-B\"{o}hm \& Rouppe van der Voort, 2009), because Gaussian functions decrease too fast with the radial distance. In addition, matching the scattered light over a small range like the range given by the observation of the transit of Mercury planet (10$\arcsec$ diameter at time of the transit, which means that the wing of the PSF can be evaluated only for a short distance of order of the radius, e.g. 5$\arcsec$) can produce only a limited knowledge of the PSF, especially beyond 5$\arcsec$ radial distance. The transit of the planet Venus of much larger diameter (60$\arcsec$) was recently used in Yeo et al. (2014) for the analysis of the PSF and for deducing the stray light correction of the smaller in diameter SDO/HMI telescope.
The stray light of the SOT is produced by the irradiation of the entrance of the telescope by the full Sun, although later the Gregorian mirror illumination is limited to the field of view (FOV); ideally the far wings of the PSF should be determined, accounting for the value of the solar radius for components before the Gregorian mirror and after, taking into account the FOV.

Here we introduce a combination of a Gaussian (to represent the core of the PSF) and a Lorentzian function (to represent the stray light at large distances) following a suggestion by Johnson (1972). It includes the far wings of the PSF of the SOT, similarly to what was done in Koutchmy, Koutchmy and Kotov (1977), to deduce the corrected extreme limb darkening functions for the near infra-red wavelengths as measured with the vacuum telescope at Sacramento Peak (Richard B. Dunn Solar Telescope). Independently, a similar PSF was also proposed by DeForest, Martens, and Wills-Davey (2009) for describing the PSF of the space-borne TRACE telescope. We use our deduced PSF to deconvolve the blue continuum images of the SOT and will discuss several new results that were straightforwardly deduced. Removing scattered light considerably improves the RMS contrast of the granulation and this is critical for fine structures such as filigree and extreme- limb features and even more for studying bright dots and structures in sunspot umbrae and penumbrae, including their dynamical behavior.

\section{Observations }
We first used images taken during the Venus transit on 6th July 2012 recorded with the broadband filter imager (BFI) of the solar optical telescope (SOT) onboard the Hinode satellite (Tsuneta et al, 2008; Suematsu et al., 2008) to determine the Point Spread Function (PSF) of the telescope, (see Fig. 1 for a selected image).  For the sake of simplicity in interpreting the photometric data, we restrict our analysis by avoiding images that were taken with filters containing contributions from chromospheric lines or from a molecular band that need a special treatment. The blue continuum images are selected because they correspond to the shortest wavelength available, presumably giving the best spatial resolution because of the favorable ratio with the given value of the aperture of the telescope. In addition, this filter gives the best temperature contrast among all continuum filters available. Similar approach was recently used by Louis et al. 2012 in their studies of umbral dots. The filter has a central wavelength at 4504.5~\textup{\AA} and a band width of 4~\textup{\AA} corresponding to a low line-blanketing effect. It is an excellent window of the continuum solar spectrum over the whole solar surface, see for ex. the recent solar atlas of Fathivavsari, Adjabshirizadeh and Koutchmy (2014).

Well focused images were selected, and each pixel corresponds to 0.05448 arcsec on the surface of the Sun. Exposure time is 205 ms, the cadence is 60 seconds and the field of view is approximately 111$\times$111 arcsec$\ ^2$.We also used a set of images taken on 1th March 2007 in blue continuum filter of a single unipolar almost symmetric sunspot images to make our movie. 125 images taken between 00:14:51 and 00:59:53 were selected in order to make it; the exposure time equals to 102 ms and the cadence varies between 6 and 60 seconds. Each pixel corresponds to 0$\arcsec$.05448 and the field of view (FOV) is approximately 56$\times$28 arcsec$\ ^2$. Furthermore, an image of this single sunspot viewed close to the disc center but with a larger FOV (223$\times$111 arcseconds) is also used. It is observed one day before on 28th February 2007 at the same wavelength and with the same pixel value (exposure time equals to 77 ms) to compute RMS contrast at disc center for magnetic and nonmagnetic regions.

Finally, we qualitatively use a CaII H line filter image (Exp time 0.205 ms with the same pixel value, 0.$\arcsec$05448) taken at the same time in order to characterize these regions in details. All images were first corrected for dark current, flat field and bias using the fg{\_}prep routine from the Solar Software.

\section{Data processing and deduction of the PSF}
The 10-arcsec-diameter Mercury disc was already used to compute and model the PSF (Mathew, Zakharov, Solanki, 2009; Wedemeyer-B\"{o}hm, 2008). The Venus disk gives a much better evaluation for the wings of the PSF because its radius is approximately 6 times bigger than the Mercury radius during the transit across the disk of the Sun. Scattered light can be seen inside the Venus disc up to a larger distance in comparison with Mercury disc (Yeo et al., 2013). In order to make the radial profile of Venus, after making several tests of the resulting signal/noise ratio (S/N), we decided to use the averaging of 30 rows (15 rows over and 15 rows under the apparent center of the planet) with respect to the diameter without significantly losing the resolution while the small curvature of the edge of Venus keeps the edge within half a pixel. Fig. 1 shows an observed image during the Venus transit and Fig. 2 shows the average radial profile of Venus using 30 rows. Because of the limited size of Venus, we further decided to use the limb of the Sun and the available part of the image of the sky to compute the far wings of the PSF taking into account the light recorded from far distance outside the disk.
\begin{figure}[t]
\begin{center}
\resizebox{7cm}{!}{\includegraphics{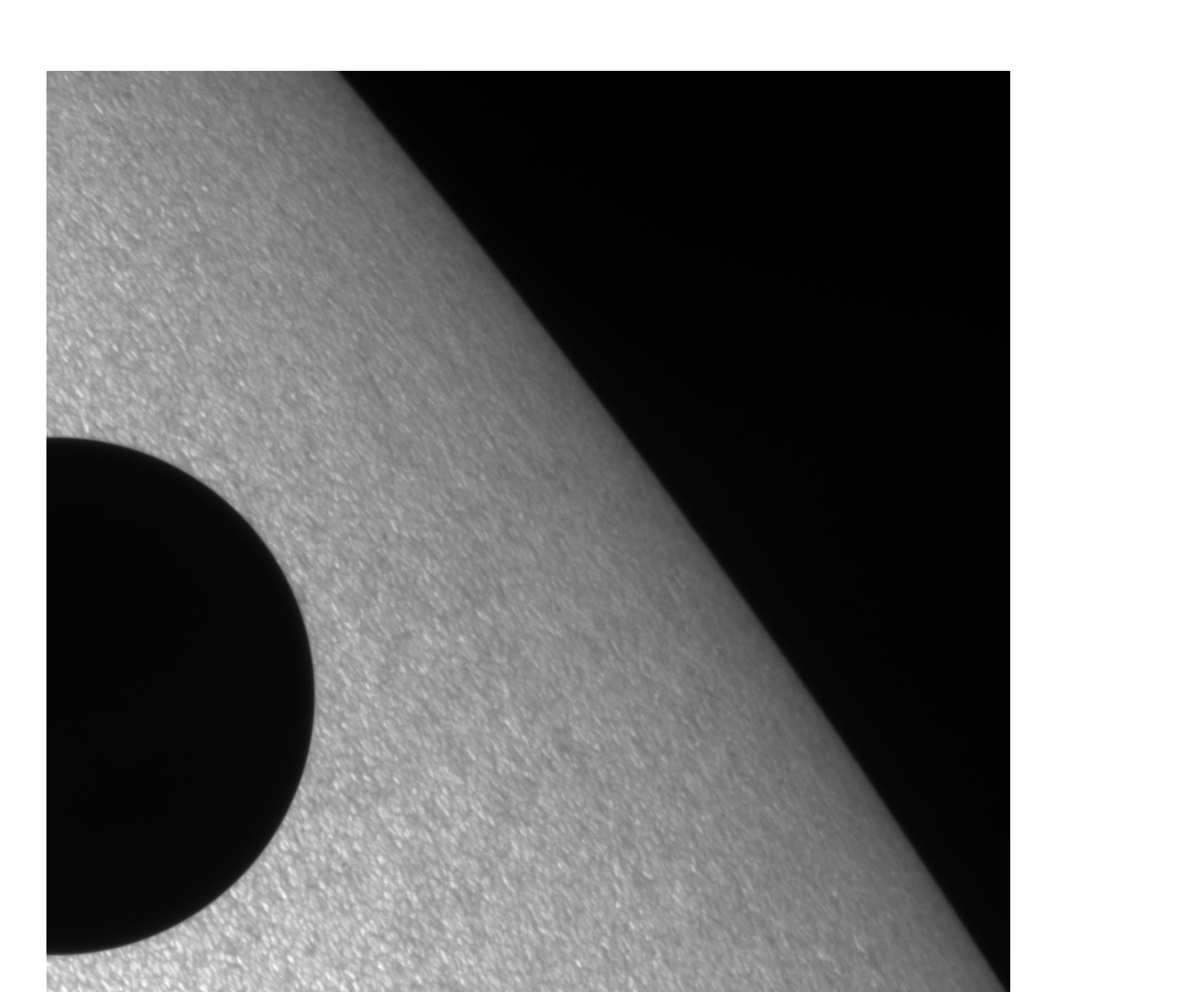}} \hfill
\caption{Observed SOT  image during the Venus transit using the blue filter. Note the limb darkening on the partial image of the Sun.}
\end{center}
\label{Fig1}
\end{figure}
\begin{figure}[t]
\resizebox{8cm}{!}{\includegraphics{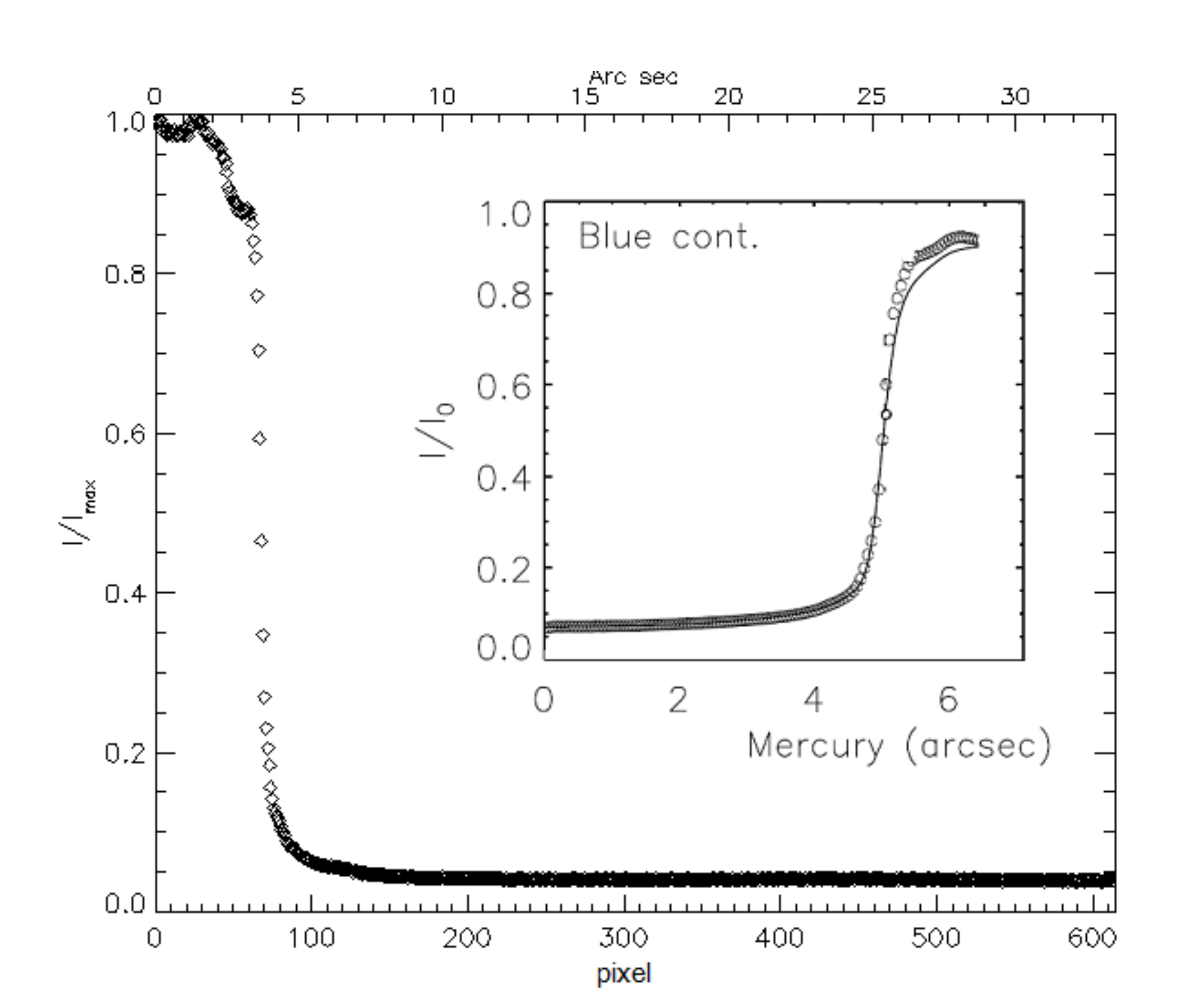}} \hfill
\caption{ Average radial profile of the limb of Venus using a 30 rows average from Fig. 1. Inserted is the cross section of the Mercury planet limb from Mathew, Zakharov, Solanki, 2009. Note the difference of scales in arcsec with the comparable level of stray light over the shadow of the planets.}
\end{figure}

In order to make a radial profile of the solar limb for deducing the smearing function, we improve the signal to noise ratio by averaging many rows, taking into account the curvature of the limb as was already done in the case of the profile of the Venus planet. Now, the curvature of the limb is much smaller and also we can go much further out from the sun where the signal is weak and the instrumental aureole is observed as in Koutchmy, Koutchmy and Kotov (1974). We can average up to approximately 400 rows without the curvature of the limb producing an effect surpassing half of one pixel. Indeed making derivative profiles of the solar limb for different numbers of rows (in pixel) parallel to a selected diameter direction (200, 400 and 600 rows), the resulting derivative profiles of the solar limb are evaluated in terms of "smoothing" when more pixels are added but taking into account the effect of curvature. It can be seen that when averaging a high number (above 400) the derivative of the limb is broadened. We use an averaging over 200 rows that corresponds to an effect due to the curvature of the solar limb of order of 0.14 pixel (half value of the sagitta). Fig. 3 shows the average limb profile of the Sun for a 200 row averaging. The ripples appearing at the left hand side are mainly due to the disk granules and heterogeneities at extreme limb (we will discuss this effect further).
\begin{figure}[t]
\resizebox{8cm}{!}{\includegraphics{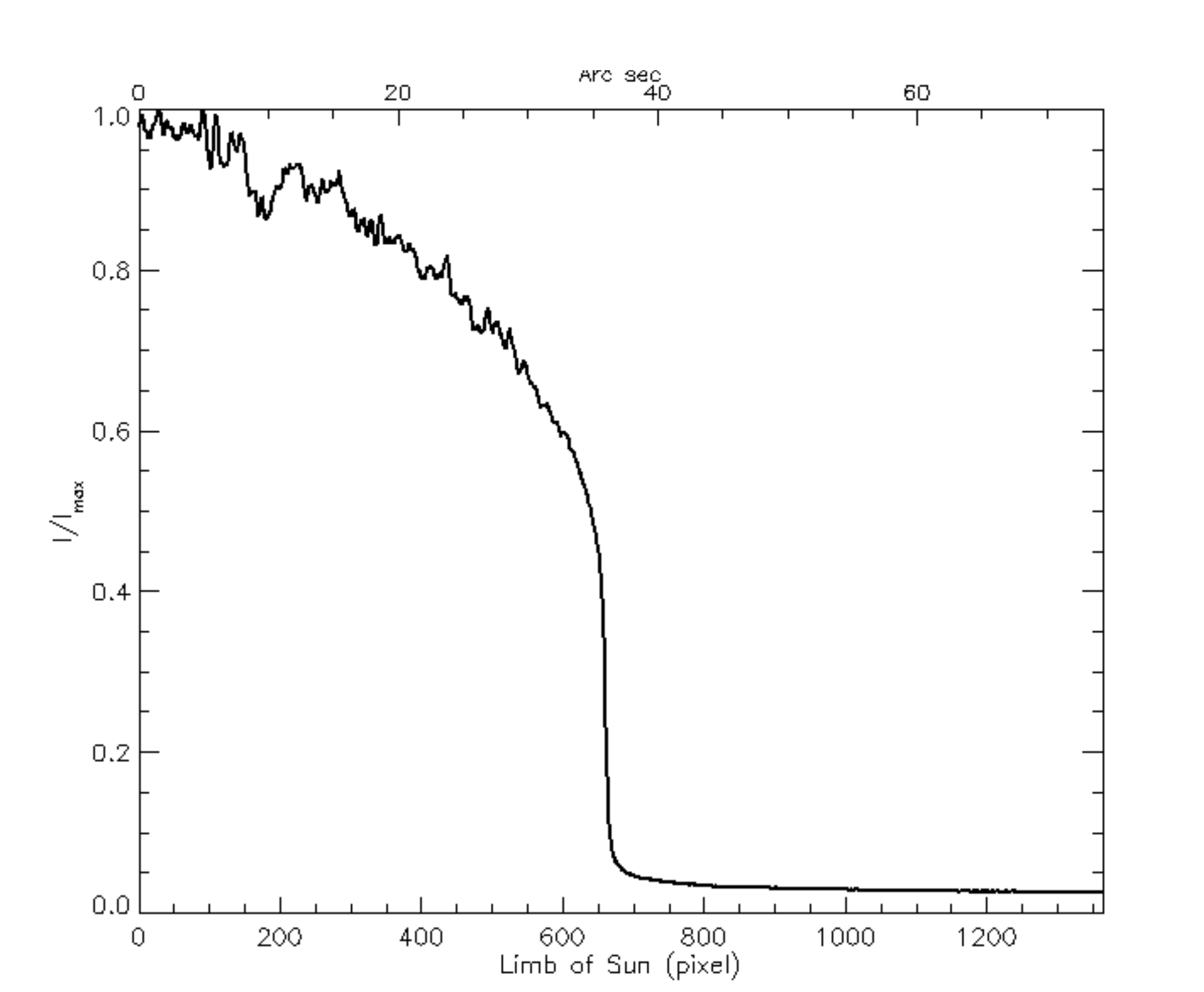}}
\caption{%
Radial profile of the limb of the sun using a 200 rows averaging.} 
\end{figure}

\begin{figure}[t]
\resizebox{8cm}{!}{\includegraphics{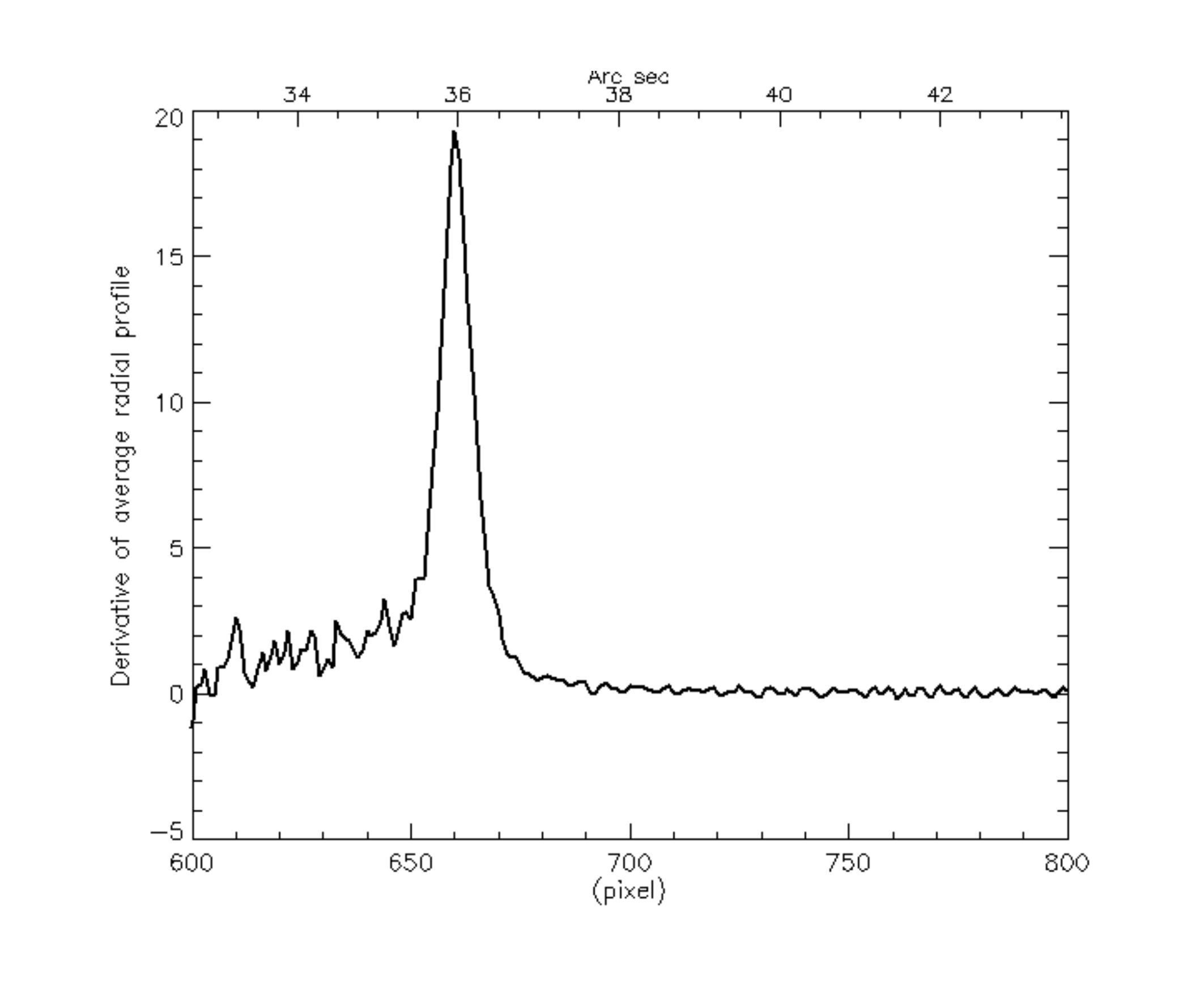}}
\caption{%
Derivative of the radial profile of the limb of the Sun shown in Fig. 3.} 
\end{figure}
Fig. 4 shows the average derivative profile of the solar limb for 200 rows. The left part of Fig. 4 is rough because of the influence of the granules but the right side does not show the influence of granules, so this curve should better be used. We then cut it through its center of gravity to make it symmetrical.  We can fit the resulted curve with an appropriate function which is the Abel Transform (AT) of the PSF, assuming the extreme limb is a step function at least at the scale we consider here, based on the theoretical expectation we have of the solar limb from a hydrostatic 1D model (Vernazza, Avrett and Loeser, 1976). In other words, the limb derivative after the inflection point is the projection of a circularly symmetric PSF along a set of parallel lines and represents the Line spread function (LSF), (see Aime, 2007 for a more elaborate discussion). The PSF could be deduced from the LSF by using an inverse Abel transform, see Bracewell (1976).
After finding the center of gravity (which is very close to abscise of the maximum intensity) we cut the peak from this point and reverse the right side to replace with the left side and have a symmetrical curve see Fig. 5. Before fitting the resulted curve, we found that it is better to rebin the data by a factor greater than one (for example 3 or 4) to reduce the role of noise during the fitting procedure and also to go further out, especially in far wings.
\begin{figure}[t]
\resizebox{8cm}{!}{\includegraphics{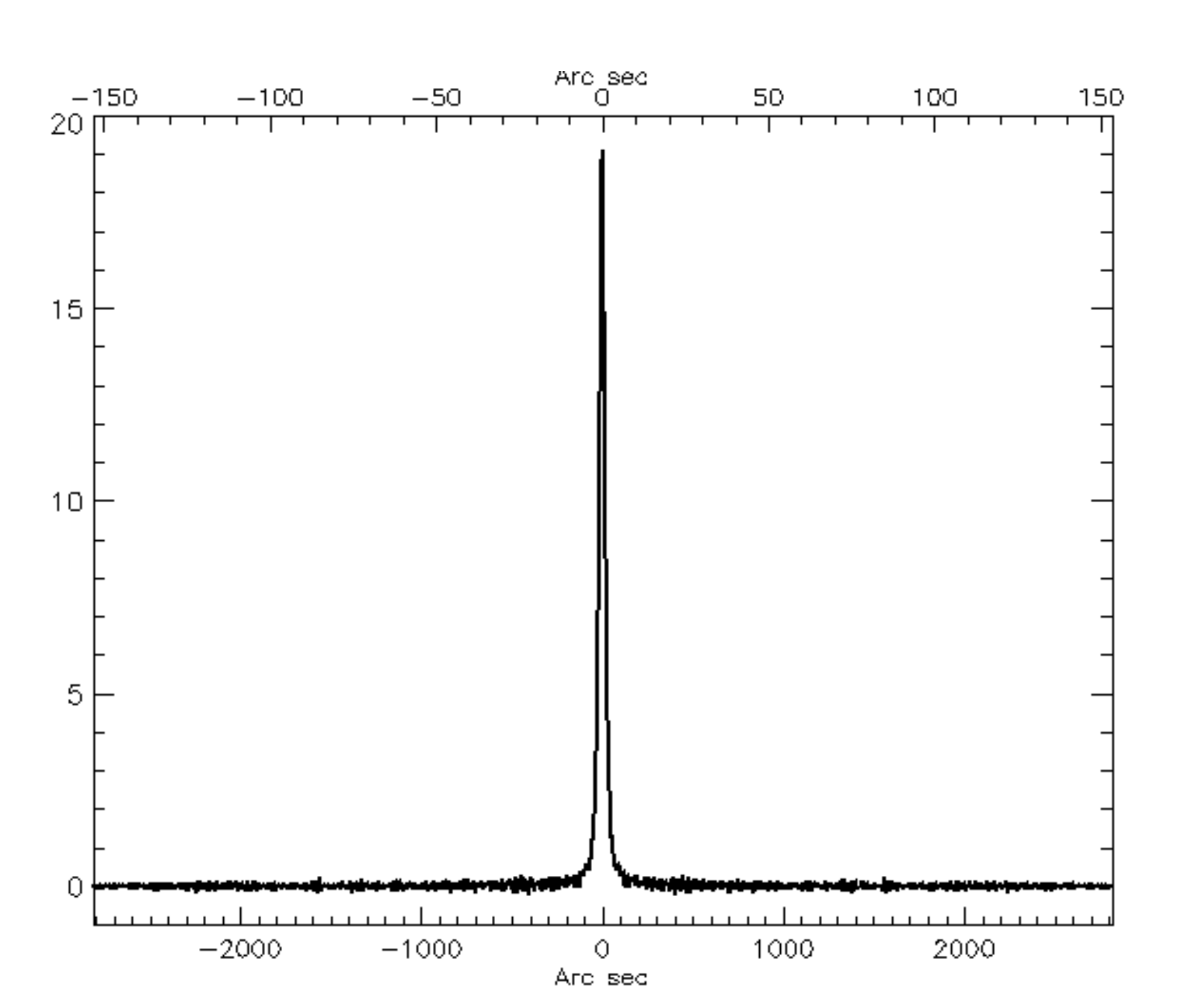}}
\caption{%
Full symmetric derivative profile of the solar limb after rebinning the data by a factor 4 to show it further out.} 
\end{figure}
We now found that the PSF can be approximated using the sum of a Lorentzian plus a Gaussian function. It means that the Lorentzian function represents the stray light that comes from far distances while the Gaussian function corresponds to short distances, including the diffraction by the entrance aperture, close to the Airy function.
The Abel transform being additive, we can fit this curve with the linear combination of the Abel transform of the Lorentzian function plus a Gaussian function. The Abel transform arises when circularly symmetric distributions in two dimensions are integrated along one dimension. The Abel transform A[f(x)] of the function f(r) is defined as

\begin{figure}[t]
\resizebox{8cm}{!}{\includegraphics{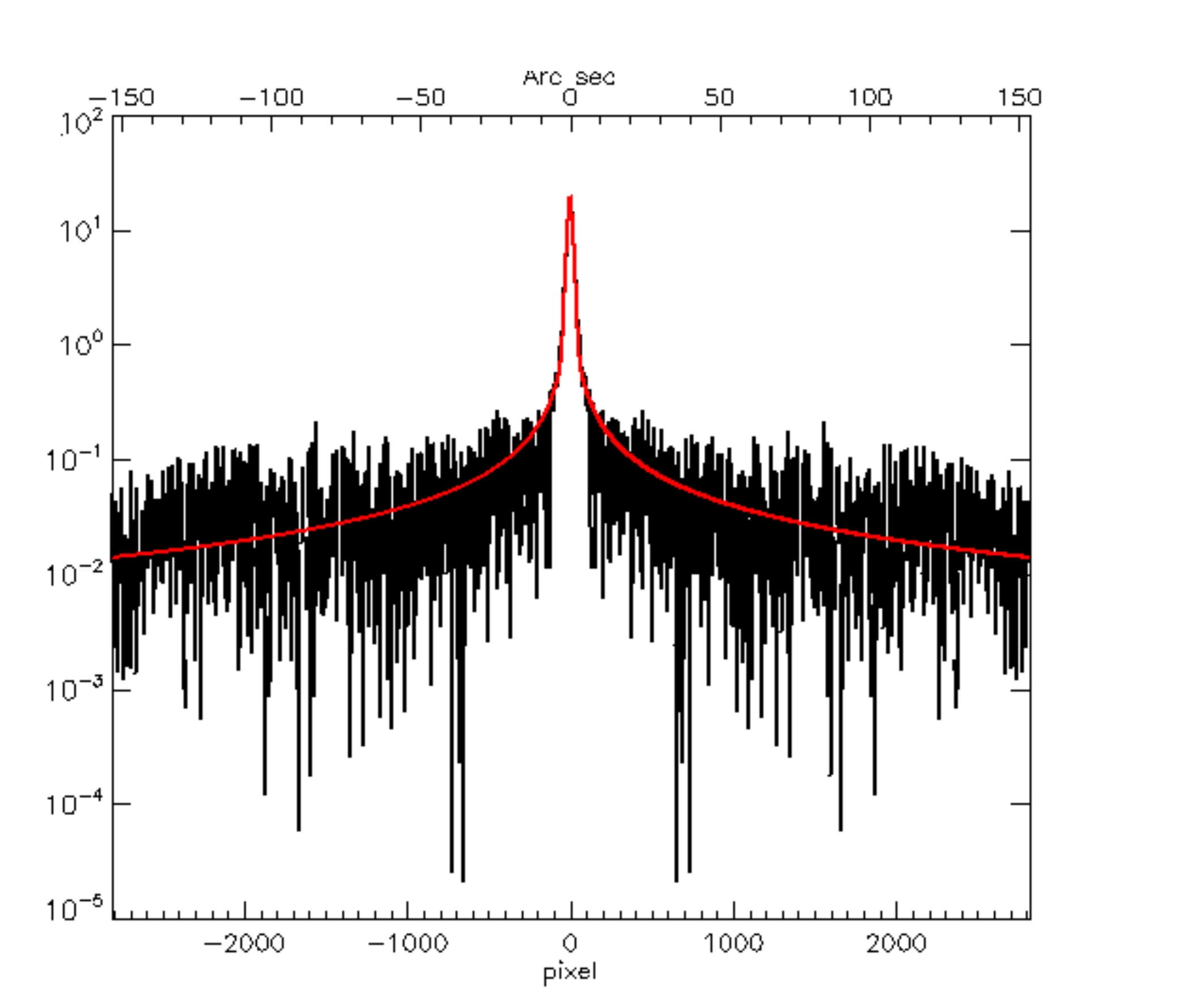}}
\caption{%
Symmetric radial profile of the solar limb after rebinning by a factor 4 (black) and the fitted line(red)} 
\end{figure}

\begin{eqnarray}
[A[f(x)]= 2\int_x^{+\infty}\frac{f(r)r\,dr}{(r^2-x^2)^{\frac{1}{2}}}
\end{eqnarray}

This transform is additive (Bracewell, 1976), it means that
\begin{eqnarray}
A[f(x)+g(x)]=A[f(x)]+A[g(x)]
\end{eqnarray}

To derive the PSF, we fit Fig. 5 with the Abel transform of the Lorentzian plus the Gaussian functions using weights pi; we call it R:
\begin{eqnarray}
R= A (p_0 L+p_3G) =p_0 A(G)+ p_3 A(L)
\end{eqnarray}
where L and G are  Lorentzian and Gaussian function, respectively. This can be written as:
\begin{eqnarray}
R=\frac{p_0\pi}{(p_1^2-(x-p_2)^2)^{\frac{1}{2}}}+ p_3\sqrt{2\pi} p_4 e^{\frac{-(x-p_2)^2}{2p_4^2}}
\end{eqnarray}

where $p_0$ and $p_3$ are coefficients (weight) that indicate the relative contributions of the Gaussian and of the Lorentzian functions in the fitting procedure,  $p_2$ is the location of maximum (Fig. 6). The result of the fit is depicted in Fig. 6, and its parameter are as follows:
\begin{eqnarray}
p_0=3.144 ,\ p_1=1.395 ,\ p_2=0 ,\ p_3=1.088 ,\ p_4=2.8
\end{eqnarray}

As discussed before, equation (1) is the Abel transform of the PSF which is a linear combination of the Lorentzian and the Gaussian functions.
\begin{eqnarray}
PSF(x)=\frac{p_0}{p_1^2-(x-p_2)^2}+ p_3 e^{\frac{-(x-p_2)^2}{2p_4^2}}
 \end{eqnarray}
Coefficients $p_0$ to $p_4$ are obtained from the fit to the observed functions. In Fig. 7 and Fig. 8 three dimensional plots of our PSF is shown; the Mathew et al. (2009) result is also shown to make a qualitative comparison. Fig. 9, and Fig. 10 show them together for a better quantitative evaluation.
\begin{figure}[t]
\resizebox{8cm}{!}{\includegraphics{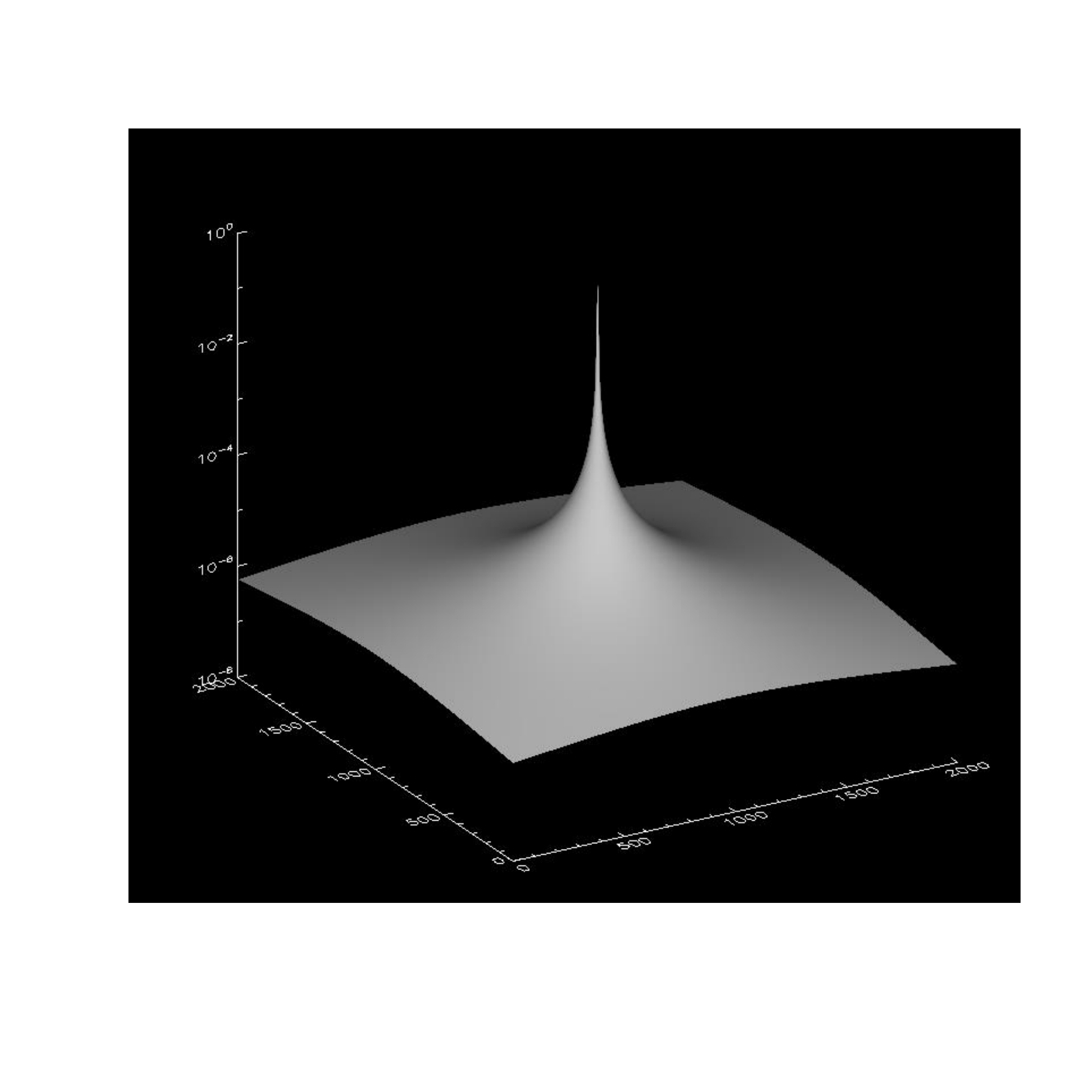}}
\caption{%
Three dimensional plot of our obtained PSF, in logarithmic scale} 
\end{figure}

\begin{figure}[t]
\resizebox{8cm}{!}{\includegraphics{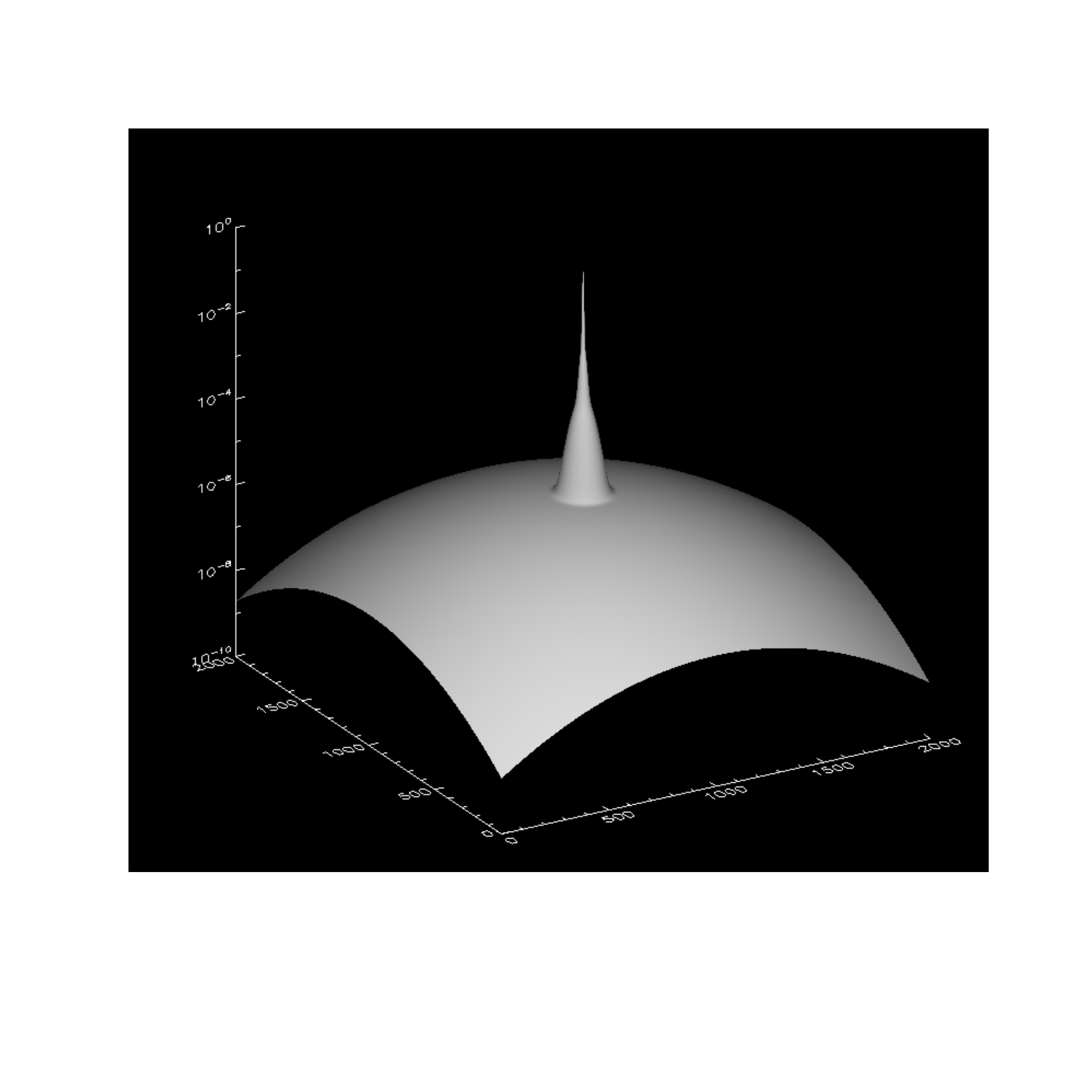}}
\caption{%
Three dimensional plot of Mathew, Zakharov, Solanki, 2009 PSF in logarithmic scale.} 
\end{figure}

\begin{figure}[t]
\resizebox{8cm}{!}{\includegraphics{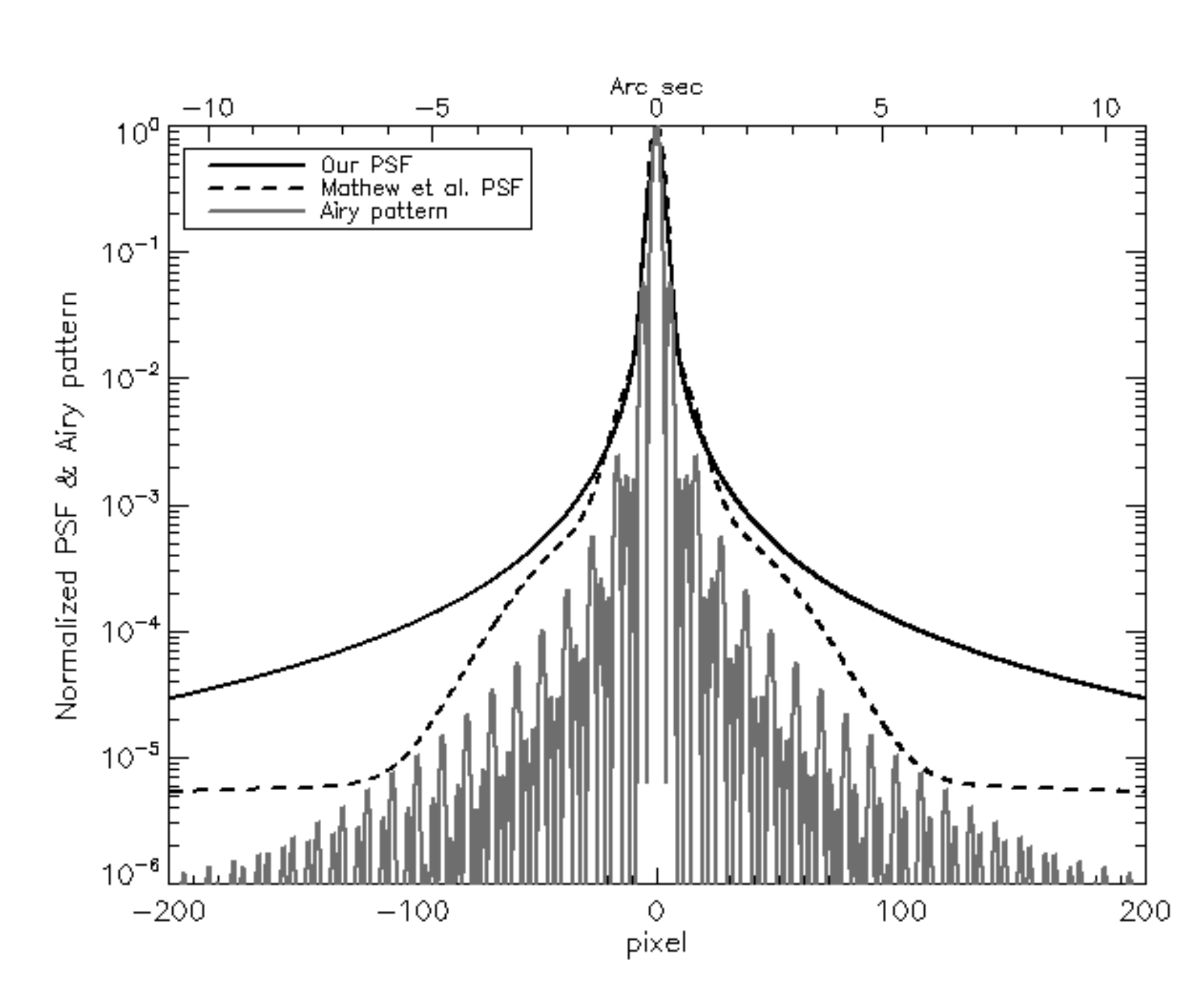}}
\caption{%
In log scale i) the theoretical Aity pattern for a full 50 cm aperture and a 450 nm wavelength (in grey), ii) the deduced effective PSF used in this paper (black full line) and iii) the one used by Mathew, Zakharov, Solanki, 2009 (dashed line)} 
\end{figure}

\begin{figure}[t]
\resizebox{8cm}{!}{\includegraphics{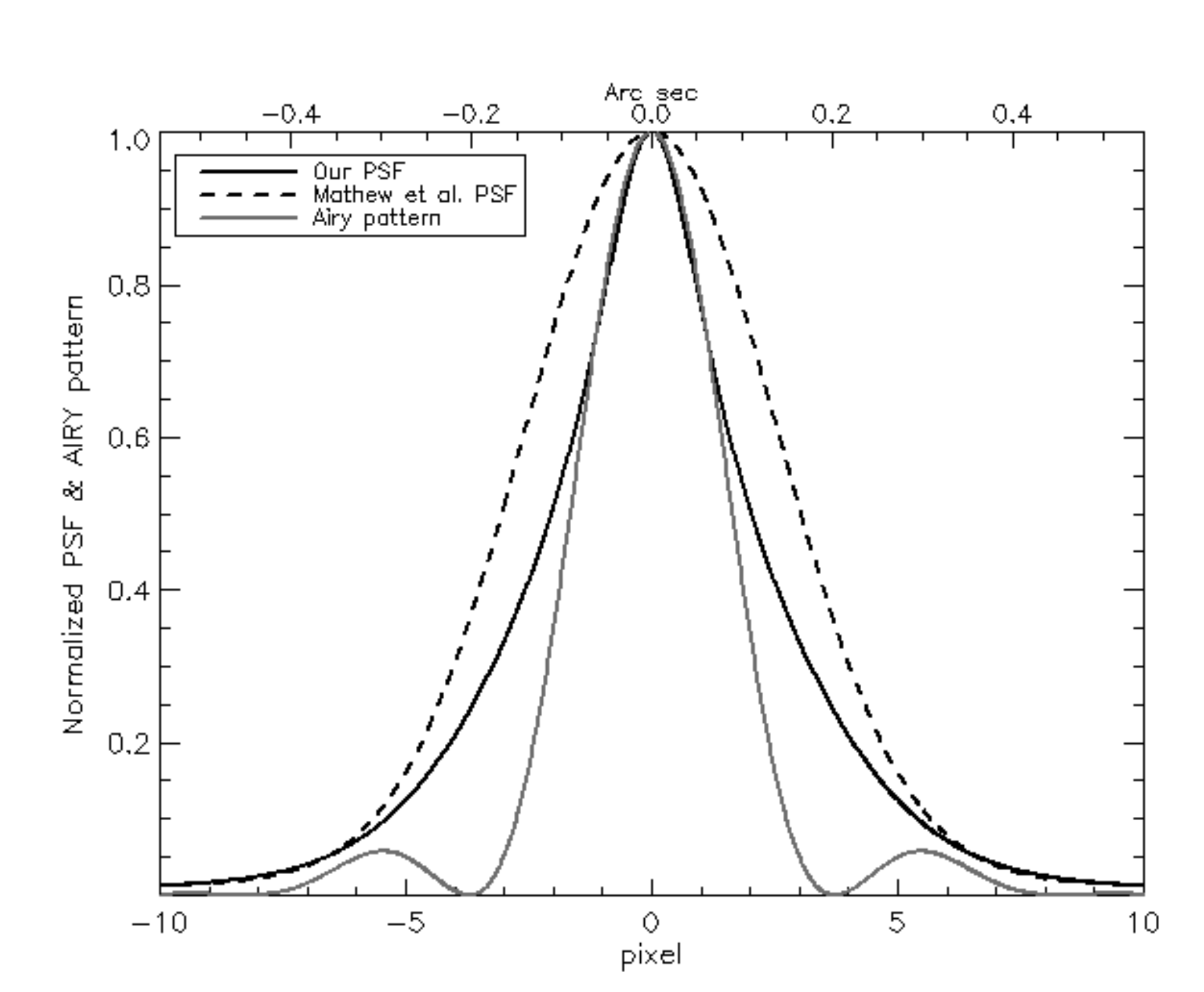}}
\caption{%
In linear scale, the Airy pattern for a fully illuminated 50 cm aperture at 450.4 nm (gray solid line), our deduced effective PSF (black solid line) and the PSF used by Mathew, Zakharov, Solanki, 2009 (dashed line).} 
\end{figure}

As it can easily be seen from Fig. 9, that the Mathew et al. (2009) PSF goes rapidly to low values after 3", so it could not fully take into account the stray light coming from moderate and great distances (when convolving using the PSF, a 920" radius Sun should in principle be considered, depending on the direction and on the position of the feature at the solar surface). Furthermore it does not have a physically and meaningful shape, since below $10^-5$ (or 6" in radial distance) its slope changes suddenly instead of decreasing slowly and continuously. They did not elaborate on the reason for this sudden change in the slope. As far as the core of the PSF is concerned, our PSF is rather narrower in the top part (Fig. 10) and this is what is expected for an aperture with a central obstruction, as was noticed from the modulation transfer function concept discussed by Wedemeyer-B$\ddot{o}$hm and Rouppe van der Voort (2009). Accordingly, we believe the PSF we propose is a good approximation of the effective PSF of the SOT, at least in the frame of the assumption of an isotropic and shift- invariant PSF. Fig. 9 and 10 show also the Airy pattern for a circular homogenously illuminated aperture of diameter d=50 cm without a central obstruction. To prove the effectiveness of our PSF in removing the stray light, we show in Fig. 11 and Fig. 13 some radial profile near the limb of the Sun, taken at a position outside of any sign of activity. Only fluctuations at the granulation scale are seen. Noticeably, no light is observed outside the limb after removing the effect of the PSF (after deconvolution, see Fig. 11 and Fig. 15) and the inflection point usually considered as showing the limit of the solar diameter, after corrections, is slightly inside the observed inflection point.

\begin{figure}[t]
\resizebox{8cm}{!}{\includegraphics{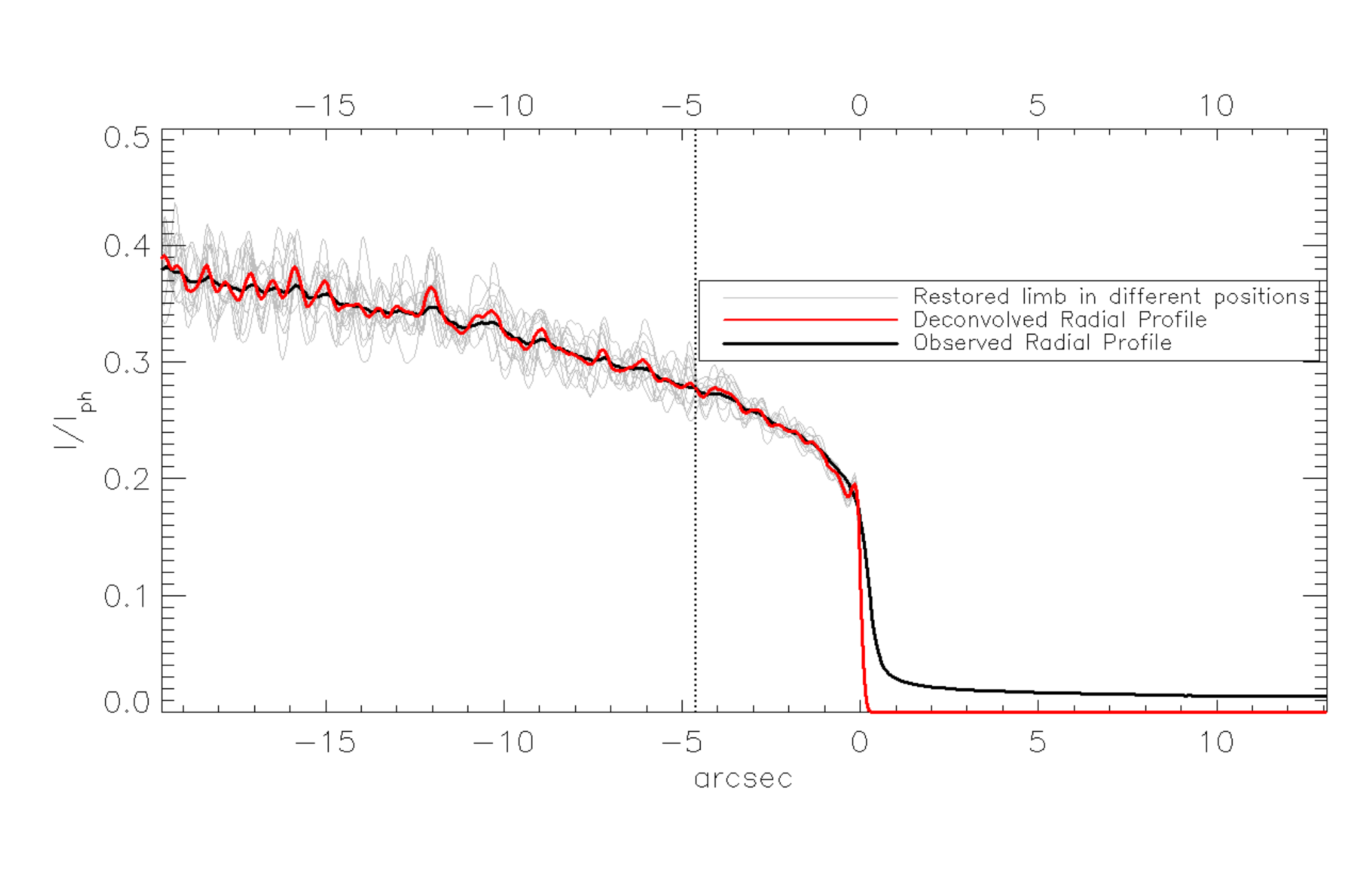}}
\caption{%
Observed (full line in black for the averaged profile) and restored limb (grey lines for narrow scans and in full red line for the averaged scan) of the Sun, vertical dashed line shows cos$\theta = $0.1.} 
\end{figure}

\section{Applications to different solar photospheric phenomena}
Restoration of images was readily performed using the method already successfully used for SOT data by Louis et al, 2012. It takes advantage of the available IDL Max-likelihood routine (Richardson, 1972; Lucy, 1974) from the Astrolib which deconvolves the observed images provided the instrumental PSF is known or assumed. This routine performs iterations based on the Maximum Likelihood solution for the restoration of a blurred image. The routine does iterations and updates the current estimate of the image by the product of the previous deconvolution and the correlation between re-convolution of the subsequent image and the instrument PSF (Richardson, 1972; Lucy, 1974; Mathew, Zakharov and Solanki, 2009; Louis et al., 2012). A parameter that we introduce when using this routine is the number of iterations to use, which depends on the quality of the original data. We used an empirical method instead of trying to discuss the relevance of the spectral distribution of the noise present in the data for optimizing the number of iterations to be used. Indeed, we are not really limited in using a large number of iterations in order to increase the contrast and, accordingly, the visibility of the features. We do not think that more than several hundreds iterations would help, as suggested by the results of the following test.

\begin{figure}[t]
\resizebox{8.2cm}{!}{\includegraphics{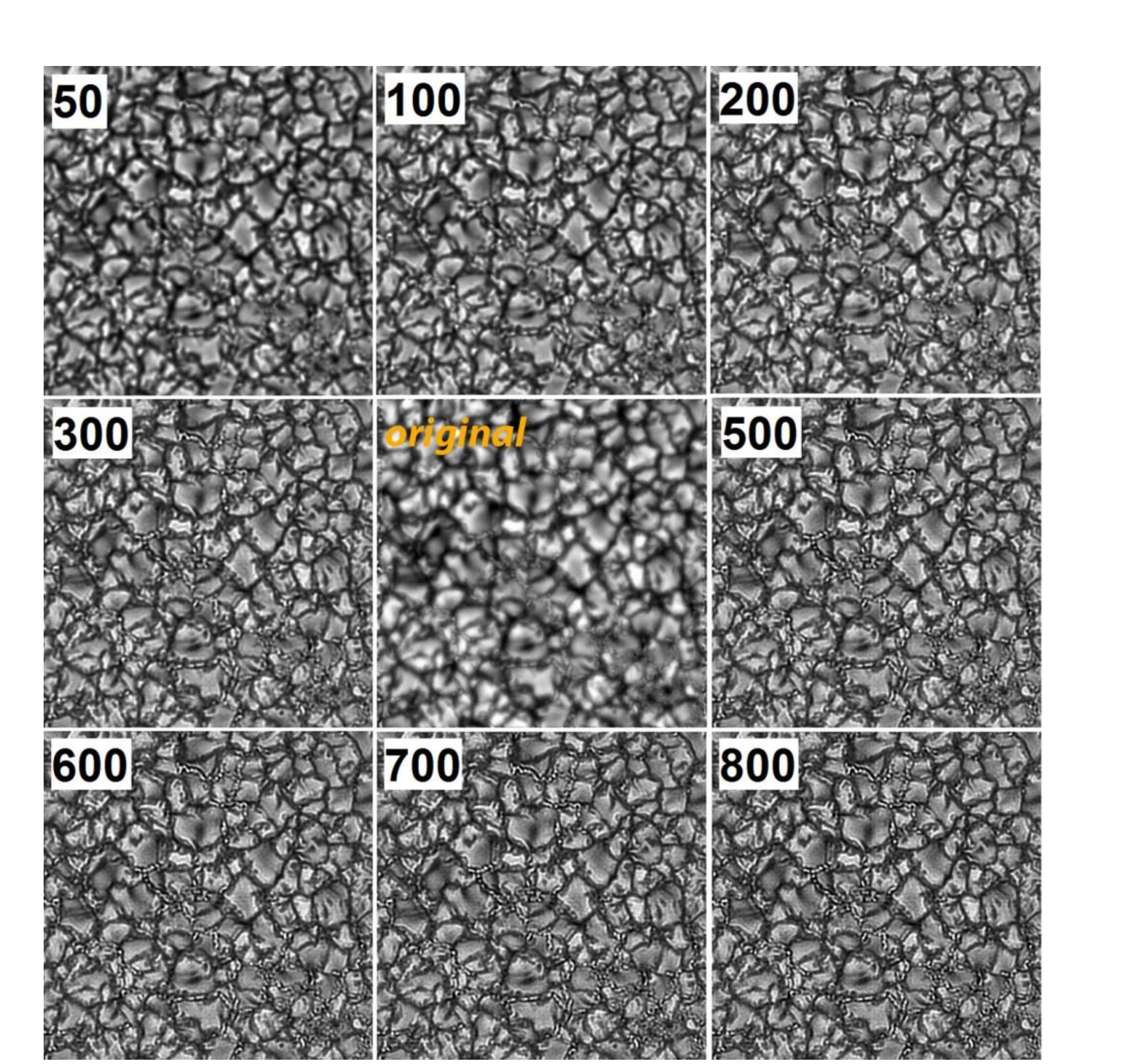}}
\caption{%
Observed (Mosaic made of the same granulation field after deconvolving using different number of iterations (see the number put at top left in each case). The original image is at the center and it is covering 25$\arcsec$$\times$25$\arcsec$area.} 
\end{figure}

A granulation field situated near the single sunspot we studied (see further), was selected to be representative of both the quiet Sun and of the magnetic granulation inserted inside (see Baudin, Molowny-Horas and Koutchmy, 1997 for a description of the magnetic (MG) and the non-magnetic granulation (NMG) that we will analyze further out). We used this test field to look how the deconvolution works by making 50, 100, 200, 300, 500, 600 ,700 and 800 iterations; Fig. 12 shows the result, to be compared with the original image inserted at the center. The impression is that the more iterations you use, the more small scale structure we see, the obvious limitation being the structures of non solar origin artificially produced by the noise. with 50 iterations we already see all features of the intergranular lanes. We could decide to stop the iteration process when spurious structures appear based on a qualitative reproducibility criteria when using consecutive in time frames. A similar procedure was used in the Koutchmy (1977) paper dealing with the deconvolution of a filigree image using the Weiner filter. This is especially true in the part where some magnetic field was measured with corresponding very fine structure called abnormal granulation (Dunn and Zirker, 1973), as at the bottom right of each box of Fig. 12. This abnormal granulation is also now called the magnetic granulation (MG), e.g. Baudin, Molowny- Horas and Koutchmy (1997). Without going too far into the details, we note that recent numerical simulations of the granulation are capable of reproducing many details of the granulation (Wedemeyer-B$\ddot{o}$hm and Rouppe van der Voort, 2009).

In order to make a more quantitative evaluation of our results, we computed the RMS of intensity fluctuations for each deconvolved image and finally, the chi-square value of the successive differences between the resulting image re-convolved with our PSF and the successive deconvolved images with different numbers of iterations:

\begin{quote}
\begin{tabular}{lll}
iterationnumber&RMSContrast&Chisquare\\
12&0.213&317.31\\
25&0.236&302.43\\
50&0.250&294.36\\
100&0.261&286.06\\
200&0.273&283.88\\
300&0.282&283.82\\
400&0.289&284.02\\
500&0.295&284.03\\
600&0.300&283.8\\
700&0.305&283.40\\
800&0.309&282.93\\

Original&0.109&\\
\end{tabular}
\end{quote}

From the examination of the chi-square values, it seems that by using 200 iterations, we reach a relative minimum in the $\chi^2$ and further, some fluctuation is seen. Accordingly, this is the maximum number of iterations adopted for the deconvolution of our images, including the case of the sunspot movie we consider below. As far as the value of the RMS of the granulation is concerned, with 50 iteration we already reach a value in agreement with Wedemeyer-B$\ddot{o}$hm and Rouppe van der Voort (2009) and the synthetic values reported by the authors. We recall that the deconvolution is done over a limited FOV; we do not take into account the whole Sun for removing the stray light which is partly reduced by the Gregory optical system used for the SOT observations. Indeed a more extended analysis could be done on this topic, but it is outside the scope of this paper, see Wedemeyer-B$\ddot{o}$hm and Rouppe van der Voort (2009) for a discussion of results coming from deeply deconvolved images when taking into account the noise.

\subsection{Study of the extreme limb}

1D models of the solar atmosphere can be tested by the behavior of their predicted center-limb variation of intensities in the continuum. The test is particularly interesting at the extreme-limb where more outer layers are concerned because line of sight integration effect. Then, the influence of the emerging magnetic field can be important, at least where the magnetic flux tubes fan out, as in the network, and are assumed to be responsible for the heating of the chromosphere. Recall that the visible limb of the Sun is typically situated at a distance of the order of 370 km from the reference height in photospheric models $\tau_5= 1$, which means in the vicinity of the temperature minimum region, below or above depending on the model atmosphere (see Rutten, 2002 for an extensive review of historical and recent models of the Solar Atmosphere). Additionally, the question of the solar diameter variation and its shape, and the extreme-limb of the Sun in its most relevant detail recently received a lot of attention (see for ex. Rozelot and Fazel 2013), not to mention the linear polarization properties at the extreme-limb which is another way to study this region. The topic is treated in so many papers including the so-called temperature bi-function effect (Rutten, 2002) that we restrict our discussion here to the analysis of measurements at 450 nm and at distances from the classical limb, defined by the inflection point, well under 1".

\begin{figure}[t]
\resizebox{8.3cm}{!}{\includegraphics{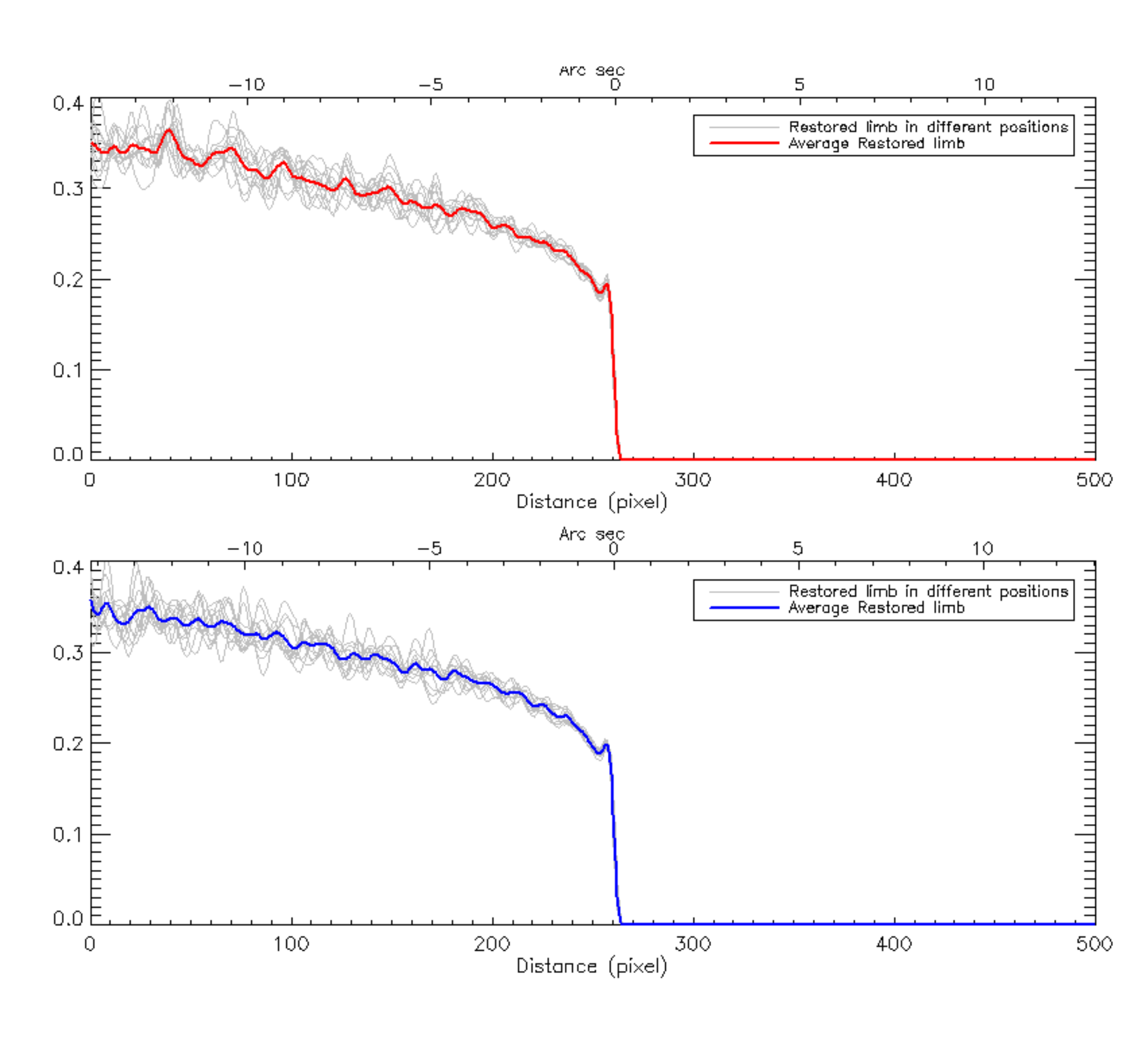}}
\caption{%
Deconvolved limb of the Sun in different positions across the limb (gray lines) as shown in Fig. 14, and their average (red and blue lines) for observations taken 6 minutes apart.} 
\end{figure}

\begin{figure}[t]
\resizebox{7cm}{!}{\includegraphics{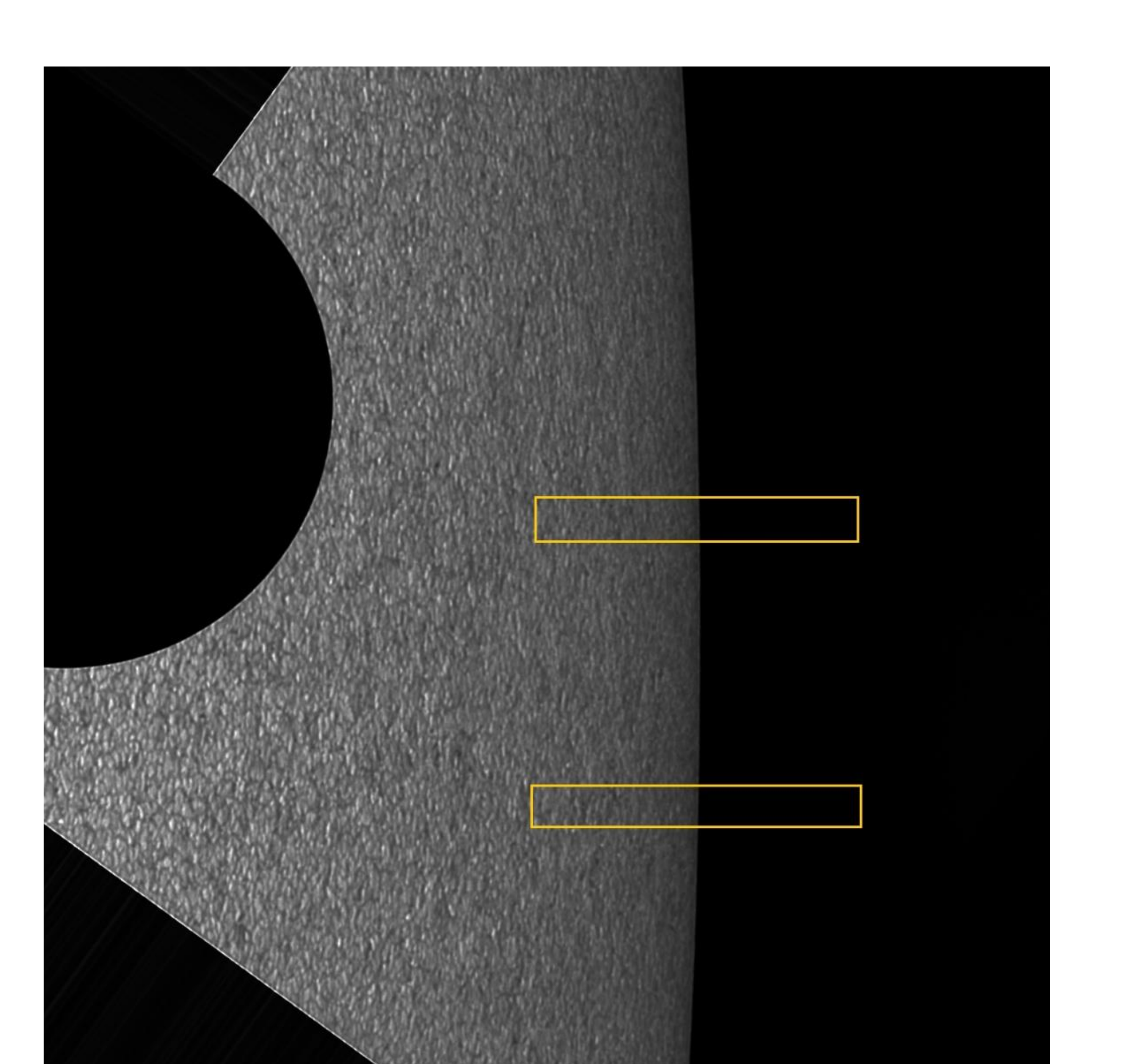}}
\caption{%
Selected image of the limb of the Sun where extreme- limb scans of Fig. 13 were obtained. The boxes show the extension of each individual scan used to get an average; the positions of the boxes correspond to the range covered by different successive averages, shown by grey lines in Fig. 13 by shifting the boxes along the limb.} 
\end{figure}

We first examine in Fig. 13 the averaged radial profiles of the limb of the Sun after deconvolution. Fluctuations are clearly apparent, up to the extreme limb, including a very small limb brightening, whose amplitude diminishes with the averaging procedure (see Fig. 14 the range of the limb positions we used). A more extended analysis is probably needed to give the precise value which possibly shows a pole equator variation. To further investigate the extreme limb, we took the derivative of the deconvolved averaged variations, see Fig. 15 and compared it with the theoretical results available from the literature, see Fig. 16. We found an excellent agreement, especially on the "width" of the limb, although inside the limb, our small extreme limb brightening is not reproduced in the former theoretical analysis of Lites, 1982 of the solar limb for different continuum windows, based on the use of the VAL classical  model (Vernazza, Avrett and Loeser, 1976) that did not yet consider the so-called "bifurcation" of temperature models (see Rutten, 2002) . The FWHM of the derivative of the theoretical limb (0.217 arcsec) is in excellent agreement with the FWHM of our deconvolved averaged limb profile (0.214 arcsec). Note that we took the Lites, (1982) data after interpolating his results to get the behavior exactly at 450 nm. Some more recent values of the limb darkening intensities were published in the literature in the frame of discussions of the solar diameter determinations with different methods, but usually they do not reach the spatial resolution of the Lites (1983) work such that we do not mention them. We recall that the Lites (1983) analysis was done to look at the so called "rugosity" of the limb, which are small-scale variations of amplitude in the height of the solar surface, presumably due to a solar granulation thought to be present in the higher layers where today we know that some reversing of the contrast occurs. The classical granulation recorded over the solar surface with high contrast could show a signature near the limb. The contrast is supposed to reverse at a height of the order of 150-200 km above $\tau_5= 1$ from models (see for ex. Gadun, 1995 and observationally, Koutchmy, 1994), at least from the analysis of images taken simultaneously  in the continuum and in photospheric lines formed near these heights. It means the reversing occurs at distance from the limb where $\tau_5$ is lower than 0.1 and the effective temperature less than 5000K, from the VAL model. Numerical 3D simulations of the solar granulation published for the last decade beautifully confirm this 1st approximation behavior. Fig. 17 is a greatly magnified part of the solar limb taken from our deconvolved image used to build Fig. 13-16 to show the fluctuations.

\begin{figure}[t]
\resizebox{8cm}{!}{\includegraphics{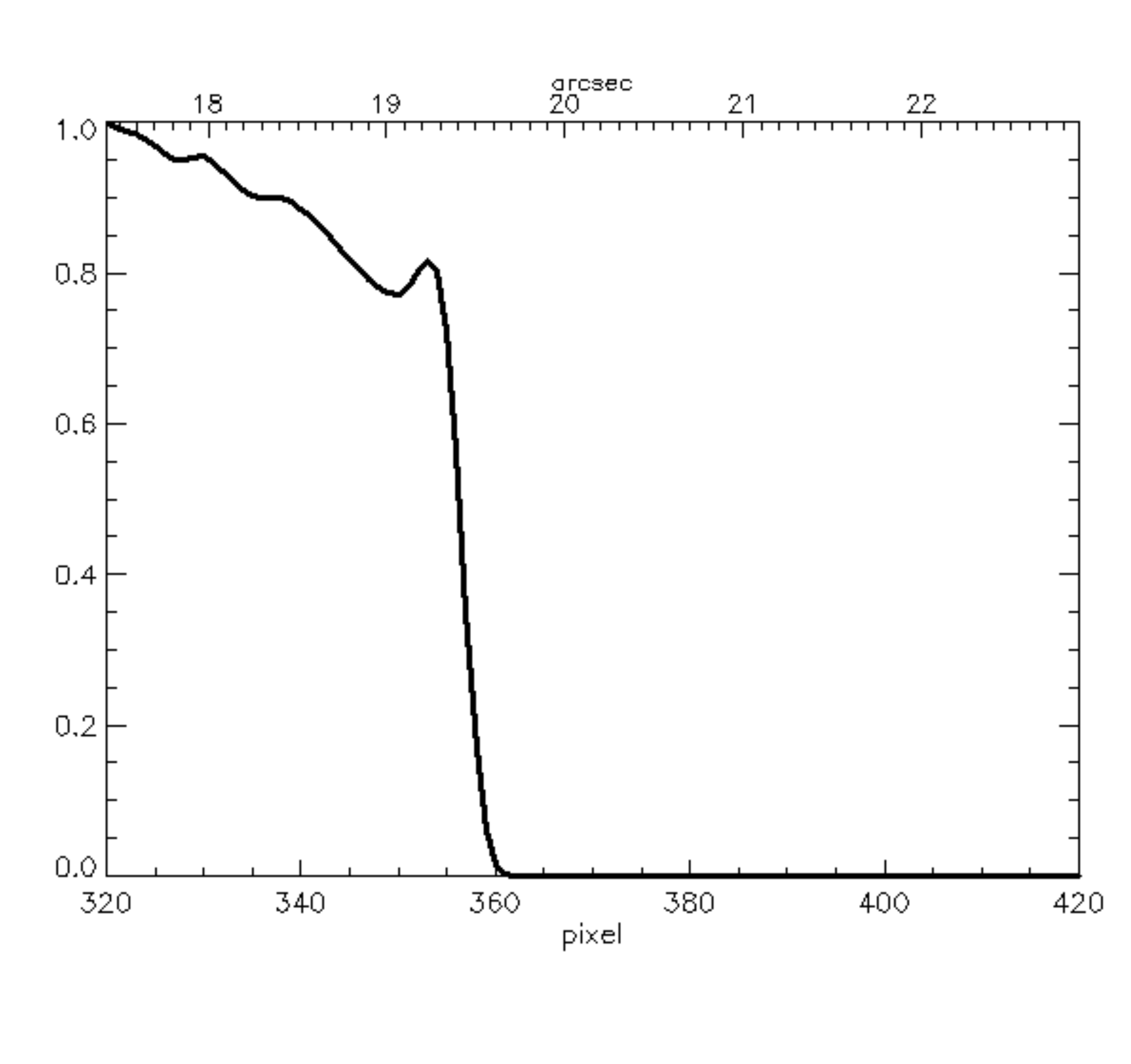}}
\caption{%
Extreme solar limb averaged variations (greatly magnified) after deconvolution in relative units, similar to the units of Fig. 15. Note that the stray light outside the limb is completely removed.} 
\end{figure}

\begin{figure}[t]
\resizebox{8cm}{!}{\includegraphics{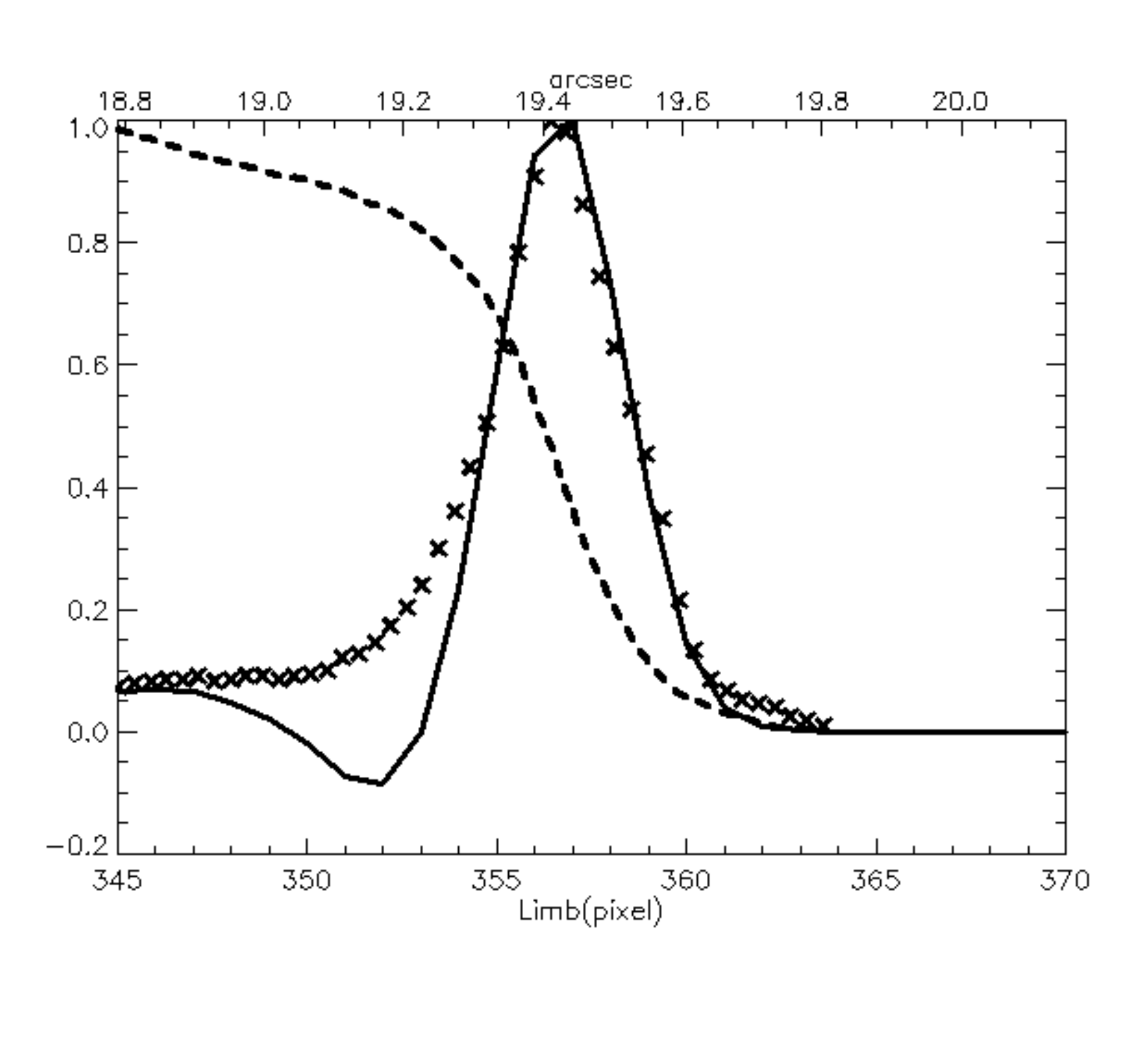}}
\caption{%
Theoretical extreme limb variation (dashed line) from Lites, 1982 and its derivative (cross symbol). Solid line shows the derivative of the deconvolved limb of Fig. 14.} 
\end{figure}

\begin{figure}[t]
\resizebox{8cm}{!}{\includegraphics{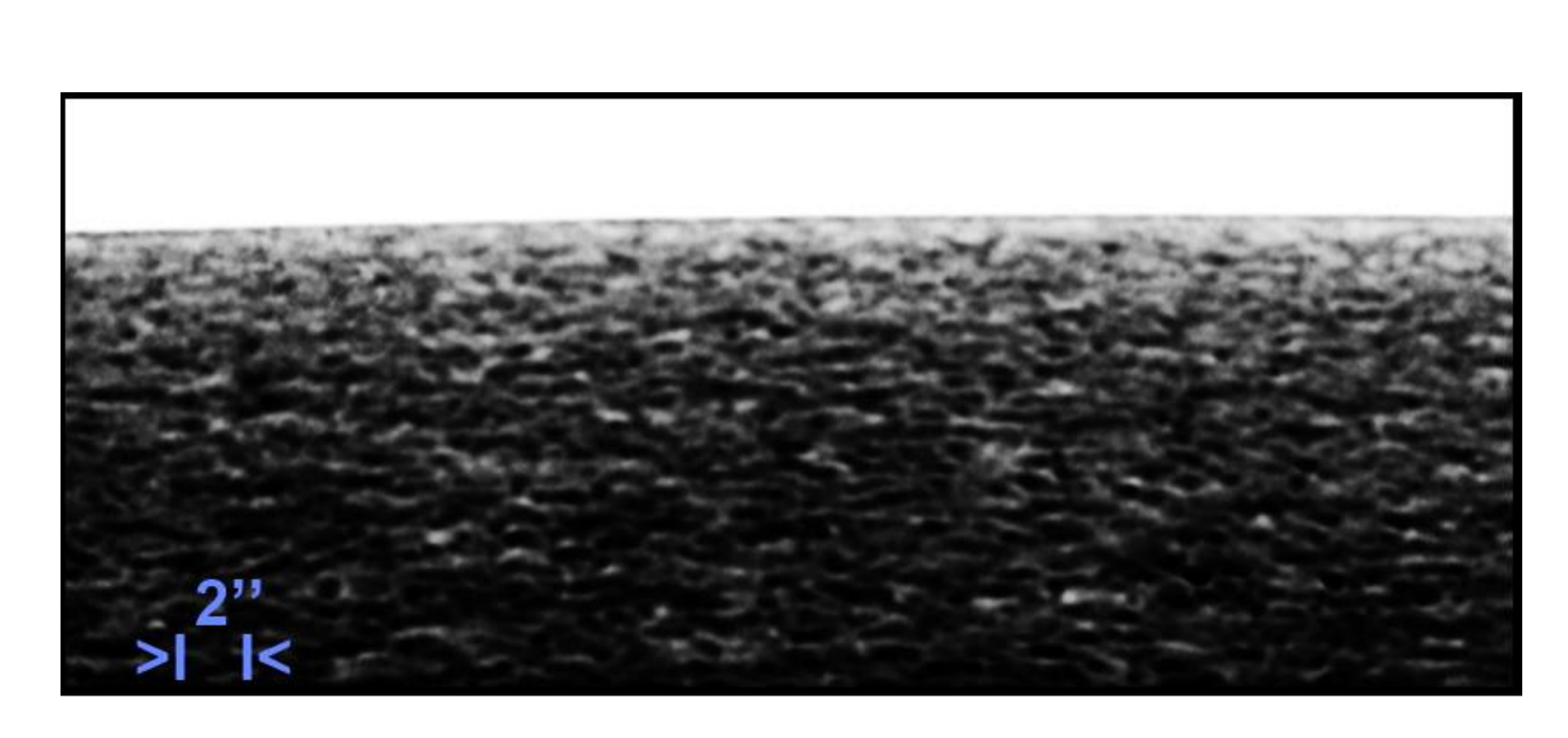}}
\caption{%
Greatly magnified image of the extreme limb variations in negative (the sky outside the bright solar disk is white). Note the structure observed up to the extreme limb and the very narrow and faint irregular limb brightening.} 
\end{figure}

Finally we note that the question of the extreme limb brightening is an old challenge for observers. The best method for observing this effect is attempted at solar total eclipses (Bazin and Koutchmy, 2013) because the level of stray light is then drastically reduced. Rosen and Poss (1982) reported a limb brightening in the last 1 arcsec from the inflection point taken as the position of the limb). We do not find an agreement with our value that shows the limb brightening much closer to the limb (see Fig. 13 and qualitatively using the greatly magnified image of the limb shown in Fig. 17.) Eclipse data based on the analysis of a recording of the eclipse light curve probably suffers from the very irregular edge of the crescent of the Moon that is difficult to take into account. Using fast flash spectra or fast slit-less spectra, a much better analysis is possible, because the edge of the Moon is imaged, see Bazin and Koutchmy (2013), from their reported flash spectra, we noticed that the rather strong "low FIP" emission line of TiII is close (450.18 nm) to the center wavelength of the blue continuum filter used in the SOT observations. This line could contribute at the extreme limb and possibly explain the observed small limb brightening.

\subsection{Analysis of the Granulation contrast over the surface of the Sun}

We applied the deconvolution procedure using our PSF on the granulation observed with the excellent focus of images taken with a pixel size of 0.05448 arcsecond during the transit of Venus. We evidently applied this deconvolution procedure in order to restore the image of Venus during its transit near the egress of the planet. The result gives us confidence in our applied correction, as all the spurious light observed on the dark disk of the planet is removed  but the study of the aureole of Venus is clearly outside the scope of this paper and will be considered later.

The deconvolution applied to the granulation of the quiet and the less quiet Sun immediately brought interesting results. The normalized RMS contrast in quiet photosphere area drastically increases after the correction is applied. The normalized RMS contrast at disk center in quiet photosphere is typically 0.11 before correction (using original observations) and it is 0.25, using 50 iterations, see Fig. 12 and further. The agreement is excellent compared to the results reported in the extended analysis of Wedemeyer-B\"{o}hm and Rouppe van der Voort, L. (2009) who developed a different method of correction, based on assumptions quite similar to ours and a PSF described with another function. It makes these results regarding the high values of the granulation contrast after correction for the PSF, rather solid and they are now a valuable test of theoretical models of granulation.

We will not extend our analysis of the granulation too far because we basically agree with the results already exposed in the exhaustive Wedemeyer-B\"{o}hm and Rouppe van der Voort (2009) paper. We prefer to invite the interested reader to study that paper that includes a theoretical part taking advantage of synthetic images produced by these authors from up to date computer simulations and models. We just add here in short some quantitative results that seem to us still interesting to show. It is in particular about the near limb granulation. Note that the study is made without taking into account the influence of p-mode intensity oscillations that indeed near the limb should not significantly change the results, from the discussion in Baudin, Molowny- Horas and Koutchmy (1997). Fig. 18 shows our normalized RMS contrast before and after deconvolution, for different distances from the limb (inflection point); typical values of the view angle $\theta$ from the normal to the local surface defined as usual with $cos\theta=\mu$ are also shown. As expected, the RMS decreases towards the limb, similarly to the limb darkening. The decrease is partly due to the foreshortening effect which makes narrower granules in the direction towards the limb. Due to the effect of dispersion of the granule intensity variations, the RMS contrast reflects the limitation of the amount of data used. however, no clear relative brightening when going to the limb is observed, noting that the part of the image we used did not show any evidence of faculae, see Fig. 17.

\begin{figure}[t]
\resizebox{8cm}{!}{\includegraphics{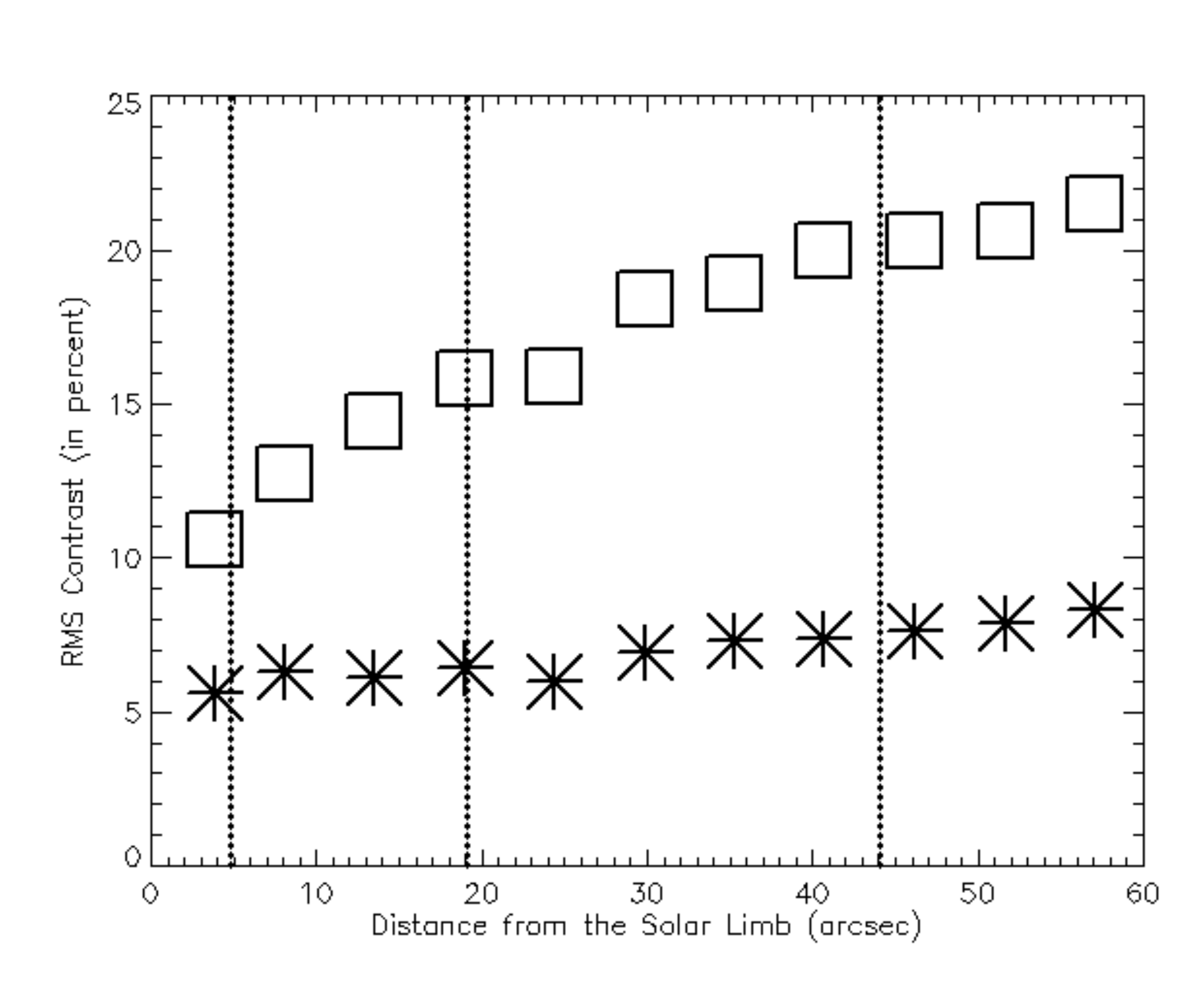}}
\caption{%
Variations of the RMS contrast near the solar limb, from both the original image and from the corrected image reflecting the behavior of the granulation contrast. Vertical dashed lines show cos $\theta= $0.1, 0.2, 0.3 respectively from left to right. } 
\end{figure}

Indeed, an interesting feature of granulation is the influence of the very small scale magnetic elements originally called filigrees that are more important in plages and also going to the limb. This magnetic granulation (MG) was carefully compared to the "non-magnetic" granulation (NMG) by Baudin, Molowny- Horas and Koutchmy 1997. Unfortunately, they used rather small spatial resolution and not corrected ground-based observations, leading to possibly limited results. Using a larger  field image from Febr. 28, 2007 coming from the observation of the same single sunspot region that we discuss after in 4.3, we produced a corrected image using our PSF. We then computed the RMS of granulation for both a non- magnetic region granulation field NMG and for a magnetic one MG, see Fig. 20. We found, after deconvolution using 200 iteration(a bigger number of iteration were used in order to increase the contrast and the visibility of small scale features): RMS (MR) = 0.297; RMS (NMG) = 0.295. The corresponding values for the original image (not corrected) RMS are 0.123 and 0.122.
Finally, we show in Fig. 19 the results of making a histogram analysis in this region taken near the solar center, see Fig. 20, to illustrate the skewness of the brightness distribution, similarly to what was discussed in Koutchmy, 1994 from the analysis of the IR (e.g. deeper) granulation observed in the opacity minimum region. The agreement is excellent with what was obtained by Wedemeyer-B\"{o}hm and Rouppe van der Voort, L. (2009) for the deconvolved normal granulation image (see their Fig.5, left part), taking into account their scale that uses the logarithmic of intensities. Making an evaluation of the difference corresponding to this bi-modal distribution which presumably corresponds to a set of bright and to a set of dark features that can be translated in temperature, we see an overall agreement with the IR results that gave a ratio 1: 0.7 of dark areas surpassing the bright areas. However now, the difference in temperature is larger because IR data were not corrected for smearing. In addition, this discussion does not include the effect of filigrees predominantly inserted in the dark lanes of the MG. It needs in a finer analysis. Fig. 20 clearly shows the difference of the field of MG and the field of NMG selected using a SOT image taken simultaneously with the H line CaII filter where plage filigrees are considerably enhanced. Fig. 19 compares what is seen in the quiet part of the image (NMG) with what is seen in a magnetic granulation (MG) region using the histogram analysis. The histograms look different around the average values of intensities (central part of the histogram); a careful examination shows a definite but really small global shift of the MG histogram towards higher values of intensities. We finally note that the skewness of the histograms revealing a bi-modal distribution, is more clearly seen for the NMG.

\begin{figure}[t]
\resizebox{8cm}{!}{\includegraphics{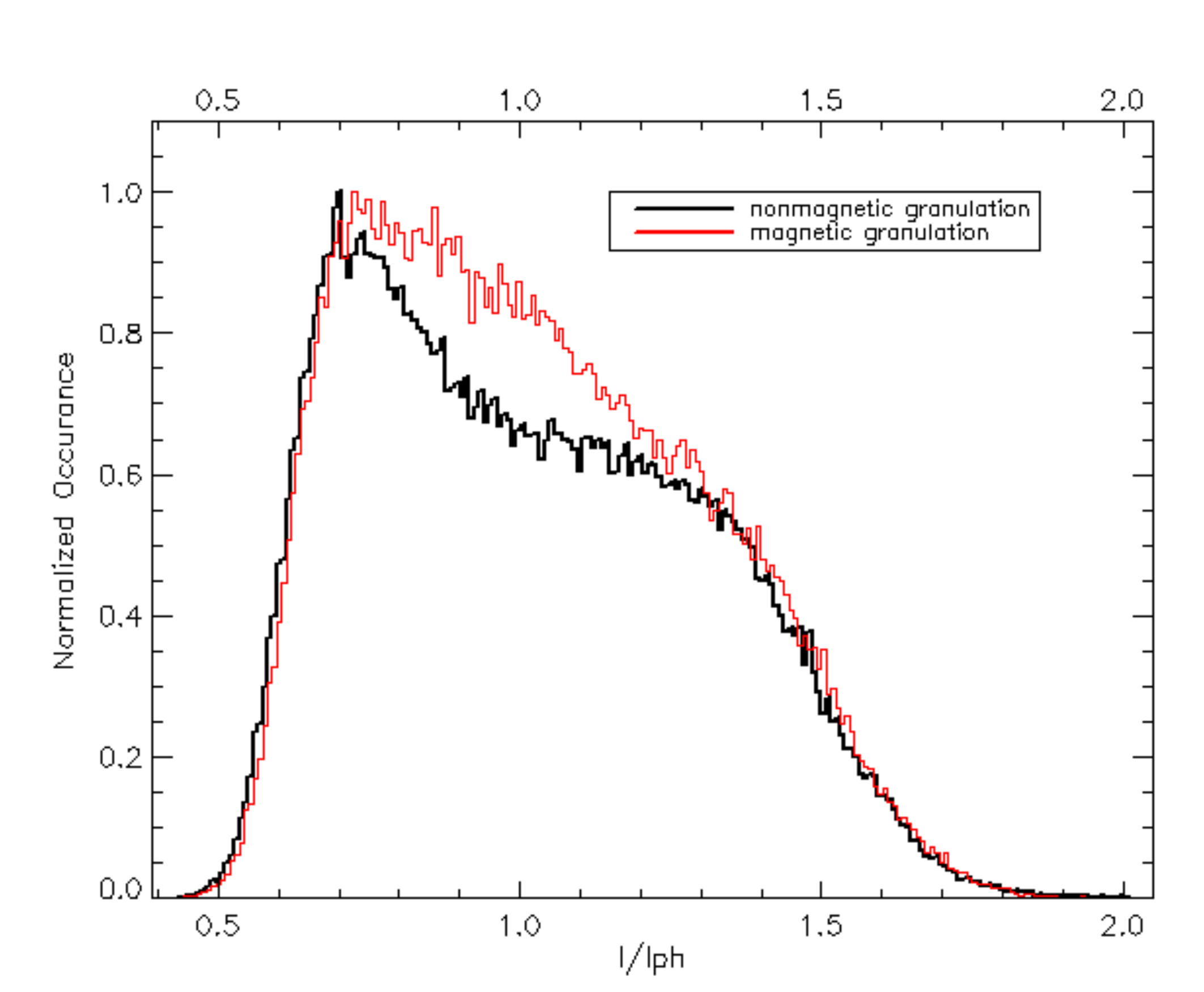}}
\caption{%
Histograms of the distribution of brightness observed over the deconvolved image of the granulation. In red, for the magnetic granulation; in black for the nonmagnetic granulation. } 
\end{figure}

\begin{figure}[t]
\resizebox{8cm}{!}{\includegraphics{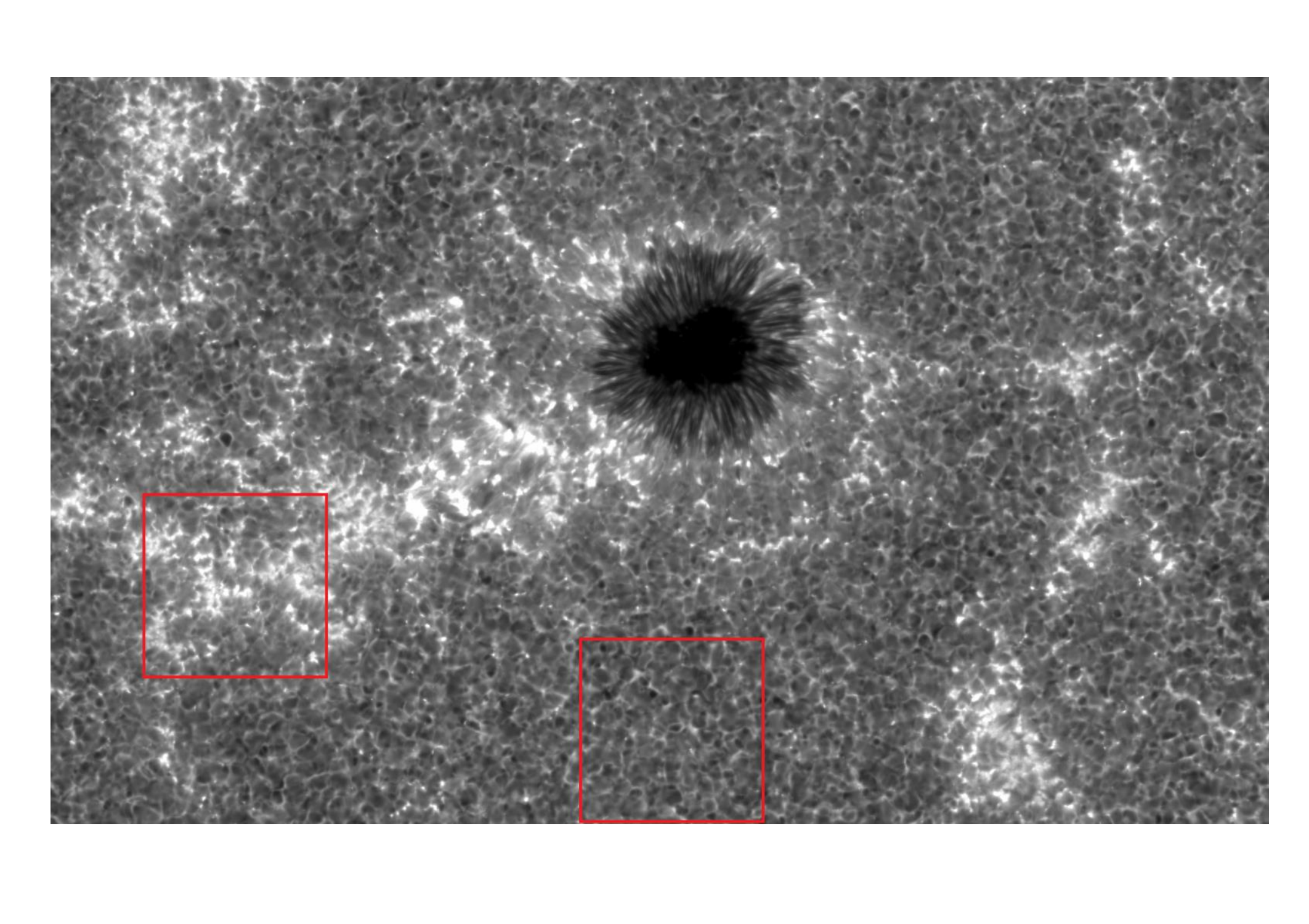}}
\caption{%
Image taken in the H line of CaII on Febr. 28, 2007 showing the selected regions for comparing the magnetic granulation MG (box at left) and the nonmagnetic granulation NMG, box at right bottom part.} 
\end{figure}

\subsection{Analysis of a single sunspot}

We now report some preliminary results corresponding to the main motivation of our study, the analysis of sunspots with the aim of understanding their origin and evolution (Solov'ev and Kirichek, 2014). Corrections needed to apply to the data play an even more important role than in the case of the quiet solar surface features, because the true umbral intensity of a sunspot core is very low. Accordingly, the stray light level in the umbra is frequently at a comparable level to the umbral brightness (see for ex. Adjabshirzadeh and Koutchmy, 1983). Moreover, the umbrae of sunspots show tiny bright features called umbral dots (UDs), of very small diameter. In the early 80s it was shown that their size is close to the diffraction limit of a 76 cm diameter telescope in the blue region (see Adjabshirzadeh, and Koutchmy, 1982-83), which means typically 0."25 (180 km), with a distribution over the umbra that makes them difficult to be definitely characterized. Indeed, clumps of UDs are seen, as well as alignments of UDs into elongated lanes. UDs are dynamical objects and their proper motion was studied (see Molowny-Horas, 1994). Qualitatively, proper motions can be demonstrated using a movie. This dynamical behavior and time variations are difficult to analyze from ground-based observations where the quality of the seeing is variable; it is much more convenient to analyze sequences with seeing-free high resolution corrected images.

Louis et al. (2012) published an extended analysis of UDs from SOT blue filtergrams processed in a way similar to our observations. However they used corrections with the PSF of Mathew, Zakharov and Solanki, (2009) which is quite different from the one we deduced here (Fig. 9 and Fig. 10), so that we anticipate differences in the results. The Louis et al. (2012) paper already describes in details the observations of March 1st 2007 that we used. Here we first provide some rather qualitative and preliminary results that apparently have not previously been seen, to our knowledge.

To evaluate the effect of the restoration method with our PSF, we apply it to the deconvolution of this "known" sunspot of March 1st (2007). Fig. 21 (top) shows an observed image and Fig. 21 (bottom) shows its deconvolved image. It can be seen from the examination of these images that the deconvolution "brings out" some umbral fine structure , such as UDs, without the need of over-exposing the picture, which means that they reach intensity comparable to the intensity of the surrounding photosphere. However, we note that the dots are observed to have a wide range of intensities and the corrected contrast depends on the number of iteration used. Regarding the peripheral UDs, their corrected intensities are even more impressive (as it was already reported in the Sobotka and Puschmann, (2009) images from the 1-m Swedish Solar Telescope reaching a 0"14 resolution). We finally decided to use 200 iteration for improving the visibility of all umbral features noting that the deconvolution does not produce any artifact effect like the undesirable ringing effect around the smallest point-like bright dots.

The contrast of penumbral filaments and the magnetic granulation around sunspot are also greatly improved. A movie was prepared from deconvolved blue continuum images taken between 00:14 and 00:59 (125 frames in 45 minutes) to see the dynamical behavior of fine structures inside umbra and penumbra. This is readily possible because of the absence of seeing in space-borne observations. Regarding the dynamics of the penumbra, the movie made of corrected frames gives some qualitative evidence of rather large scale emerging penumbral diffuse dark radial magnetic strands moving together with elongated bright quasi-radial filaments with evidence of twist and dark lanes inside. Simultaneously, there are bright features moving inwards with detachments making the peripheral UDs. At smaller scale, a more careful analysis, possibly with additional spectroscopic material, should be done before discussing the physics of the penumbra. The movie however shows excellent examples of the inwards motion of bright rather elongated elements towards the periphery of the umbra produces peripheral UDs with a definite excess of brightening, suggesting that nonlinear processes producing a shock could occur at the umbral boundary where the magnetic field sharply changes of direction. In addition, the penetration process into the umbra continues with evidence of splitting. These phenomena will be addressed more quantitatively in the future.

\begin{figure}[t]
\resizebox{8.4cm}{!}{\includegraphics{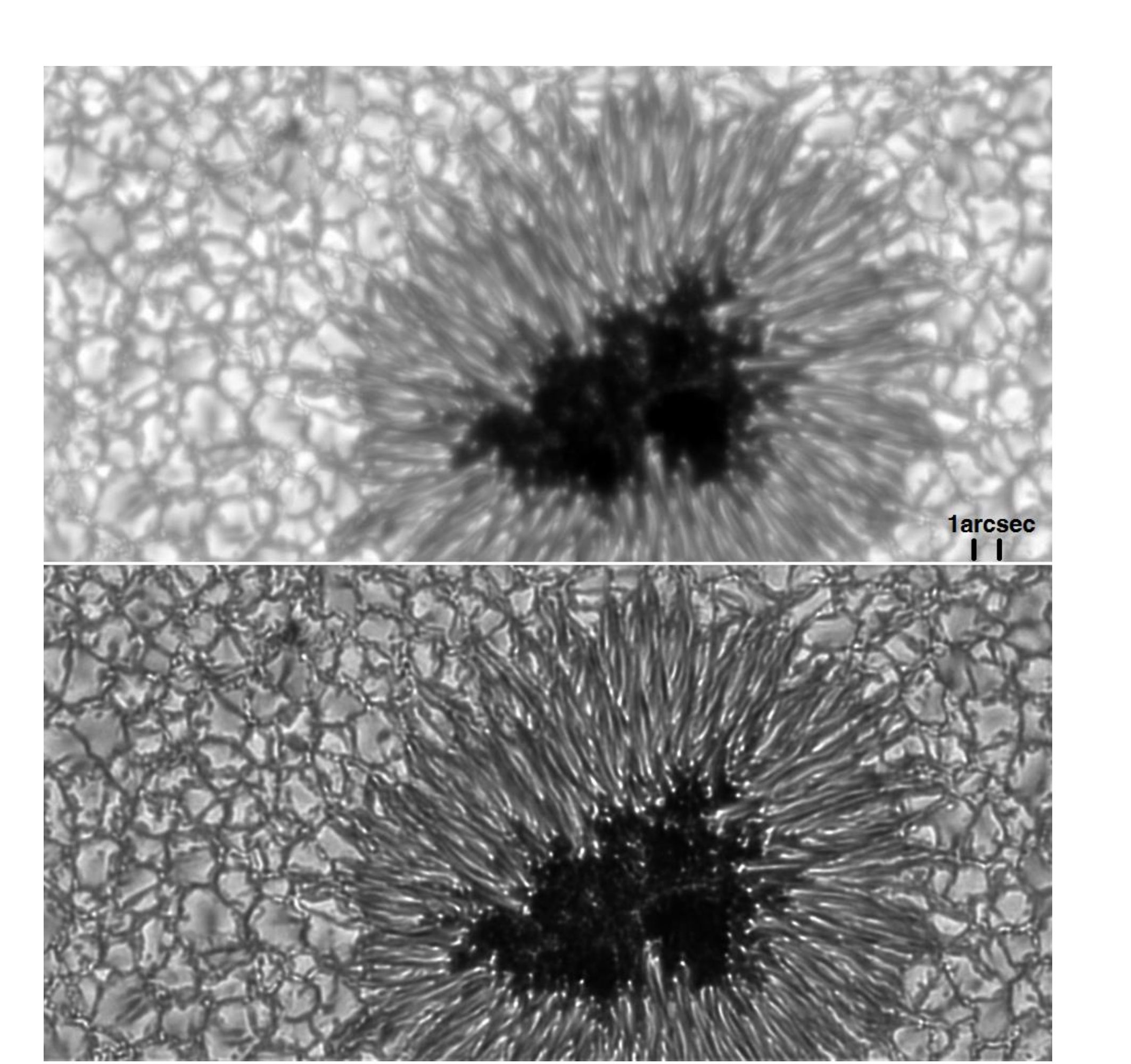}}
\caption{%
Observed SOT image of the single sunspot of 2007 March 1 (top) and the deconvolved image } 
\end{figure}

A last result we want to show concerns the most critical part of the sunspot, from the point of view of the origin of the strong magnetic flux emerging from the deep layers responsible for a large part of the activity of the Sun which is concentrated in the core. Fig. 22 is a display made to show the structure of the core, with UDs visible inside, including i) an alignment pattern of UDs partly making a very weak bridge, with a dark lane inside, not far from the central part of the core; their morphology reminds us of what is observed in the classical light bridges of sunspots (Secchi, 1870) but here, we deal with a very low intensity feature; ii) an area in the core close to the weak bridge, where, surprisingly, no UDs are present, making a sort of flat area of approximately 3$\arcsec$$\times$2$\arcsec$ surrounded by several elongated UDs, corresponding to the darkest part of the umbra. We have not found in the literature a description of a similar feature in the umbrae of sunspots. These peculiarities cannot be detected on the uncorrected images. We believe that further analysis is needed to confirm their presence before speculating on the physical significance of this finding. Hopefully, such confirmation will come from a much deeper analysis of our movie material permitting to tackle very small scale dynamical phenomena.

\begin{figure}[t]
\resizebox{8.2cm}{!}{\includegraphics{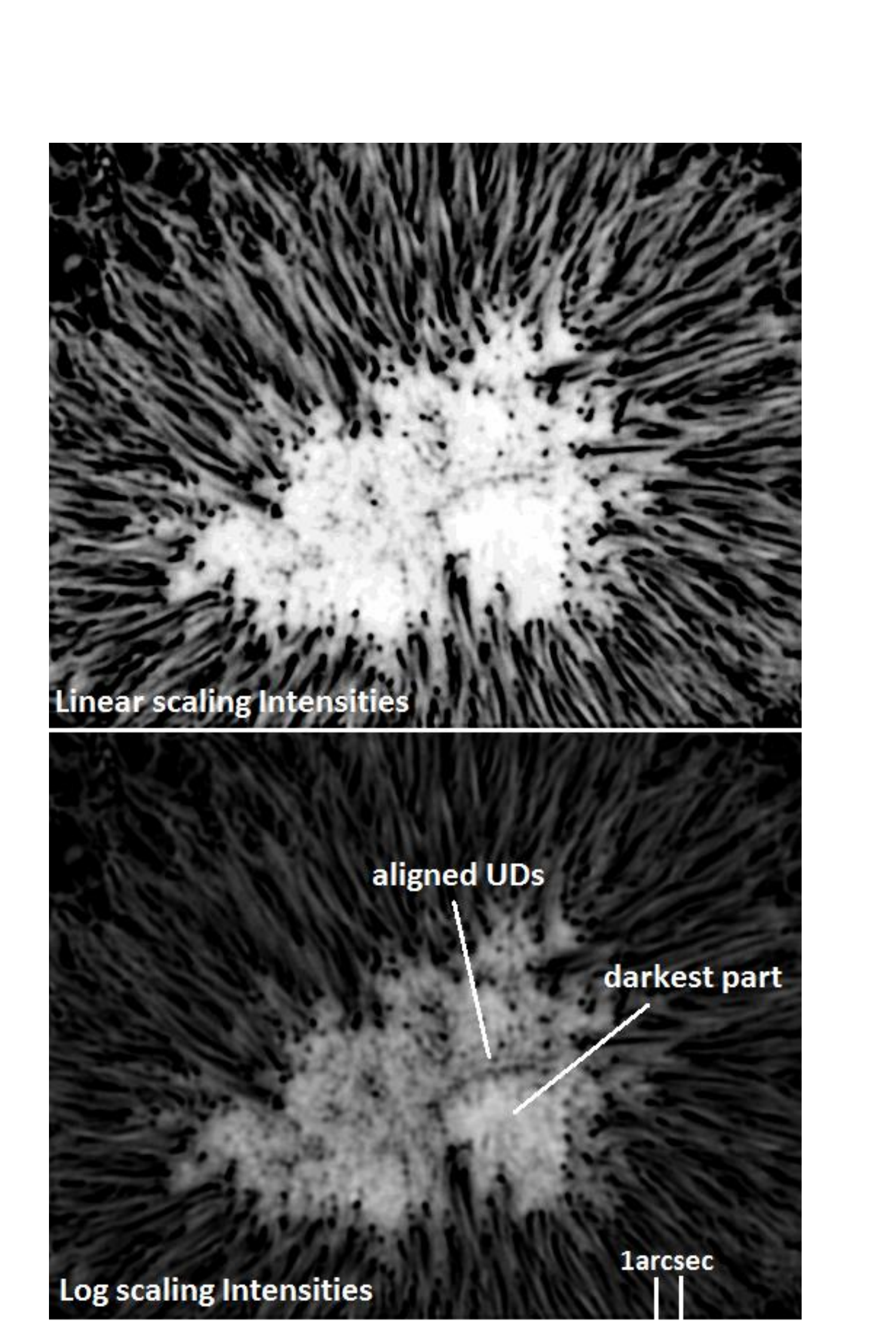}}
\caption{%
Negative display in both linear and logarithmic scales of intensity modulations of the core of the sunspot showing peculiar features of the umbra. } 
\end{figure}

\section{Conclusion}

The blue continuum high resolution images of the SOT of the Hinode mission were corrected for stray light based on several observations, including the use of images of the transit of planet Venus and the case of the extreme limb 1D variation. A combination of Gaussian and Lorentzian Functions were used to construct a new telescope PSF. This combination of functions well represents the diffracted light that comes from narrow angles (core of PSF) and the stray light that comes from large angles (far wings of the PSF).

The subsequent deconvolution using iterations of images greatly improves the contrast of any image of the solar surface. In the case of the aureole of the planet Venus, an impressive gain is obtained for looking at its real intensity variations, it removes all the spurious light from the dark disk of the planet and it gave us confidence in our method of restoration.

The normalized RMS contrast of the quiet photosphere is increased by more than a factor 2 and some statistical properties of the granulation, inside and outside magnetic regions such as a plage with filigrees inserted, are deduced. Furthermore the corrected images show a new result regarding the extreme limb of the Sun, including a faint limb brightening at 0"2 of the edge of the Sun where indeed elongated along the limb irregularities are still observed. No interpretation is offered but it suggests that the simultaneous analysis of images obtained with different filters could help.

Finally, deconvolved images of a sunspot also show some new dynamical penumbral structures and especially some new umbral structures much more clearly, such as aligned umbral dots (UDs) seemingly making the start of an umbral bridge. The core of the umbra also shows an area where UDs are absent, which seems difficult to reconcile with an assumed vigorous magneto-convection present permanently inside the umbra. The deduced movie is proposed as additional material for the future analysis of small scale dynamical phenomena. It needs careful study, including correlation analysis of successive images with short time interval that we intend to describe in a forthcoming paper. We believe this is now possible because i) images are free of any turbulent effect, ii) their photometric properties are excellent and iii) the proposed correction based on the use of a rather simple PSF makes the deduced results solid, confirming that we deal with wonderful material.

\acknowledgments Hinode is a Japanese mission developed and launched by ISAS/JAXA, collaborating with NAOJ as a domestic partner, NASA and STFC (UK) as international partners. Scientific operation of the Hinode mission is conducted by the Hinode science team organized at ISAS/JAXA. This team mainly consists of scientists from institutes in the partner countries. Support for the post-launch operation is provided by JAXA and NAOJ (Japan), STFC (U.K.), NASA, ESA, and NSC (Norway). The work of H.G. is supported by an award of the French Embassy in Teheran. We thank P-A. Lhote, L. Vigroux and F. Bernardeau for their constant interest and support. We thank Hassan Fathivavsari, Jean-Claude Vial, Frederic Baudin and specially Leon Golub for providing a critical reading of the paper. Finally we want to thank our referee for pointing out excellent remarks that permitted to significantly improve the paper.

\makeatletter
\let\clear@thebibliography@page=\relax
\makeatother

\end{document}